\documentclass[aps,twocolumn,floatfix]{revtex4-2}
\usepackage{bm}
\usepackage{dcolumn}
\usepackage{graphicx}
\usepackage{amsmath}
\usepackage{amssymb} 
\usepackage{amsfonts}
\usepackage{float}
\usepackage[capitalise]{cleveref}
\usepackage{xcolor}

\usepackage{array} 

\newcommand{\RNum}[1]{\uppercase\expandafter{\romannumeral #1\relax}}

\begin{document}
\title{Magnetoplasmons in $N$-layer structures} 
\author{Jinu Park$^{1}$}
\thanks{These authors contributed equally to this work.}
\author{Taehun Kim$^{1}$}
\thanks{These authors contributed equally to this work.}
\author{E. H. Hwang$^{2}$}
\author{Hongki Min$^{1}$}
\email{hmin@snu.ac.kr}
\affiliation{$^{1}$ Department of Physics and Astronomy, Seoul National University, Seoul 08826, Korea}
\affiliation{$^{2}$ SKKU Advanced Institute of Nanotechnology and Department of Nano Engineering, Sungkyunkwan University, Suwon 16419, Korea}
\date{\today}

\begin{abstract}
{
We provide a systematic framework to investigate the magnetoplasmons of multilayer two-dimensional electron systems by using the Kac--Murdock--Szeg\H{o} (KMS) Toeplitz matrix to consider interlayer Coulomb interactions.
In the absence of interlayer tunneling, we show that the single-layer magnetoplasmon branch splits into $N$ collective modes---one in-phase mode and $N-1$ out-of-phase modes---and derive their asymptotic behaviors in the long-wavelength limit, as well as in the limit of large layer separation and strong magnetic fields.
When interlayer tunneling is present, we clarify the magnetoplasmon dispersion, both qualitatively and quantitatively, by identifying the magnetoplasmon mode associated with each interband transition, as well as tunneling magnetoplasmons arising from interband transitions with the same Landau level index.
Our study presents the hybridization between the modes governed by underlying symmetries, along with an enhanced tunneling magnetoplasmon gap exceeding the associated interband gap.
The KMS-based analytic formalism thus provides a comprehensive physical understanding of magnetoplasmons in multilayer structures.
}

\end{abstract}

\maketitle

\section{Introduction}
Since the realization of two-dimensional (2D) materials such as graphene, transition-metal dichalcogenides, and their heterostructures, multilayer systems have opened up a new area of 2D physics.
Recent advances in experimental techniques such as molecular beam epitaxy and exfoliation~\cite{Cho1975, Novoselov2004, Coleman2011} have enabled the fabrication of high-quality layered samples including graphene, quasi-2D materials, and GaAs quantum wells ~\cite{Dingle1974, Novoselov2004, Radisavljevic2011, Chhowalla2013, Lin2018a, Trang2021, Lei2022}.
Controlling interlayer interactions through
stacking order~\cite{Profumo2010, Jang2015},
twist angles~\cite{Liu2014, Shin2023},
and electrostatic gating~\cite{Gorbachev2012, Burg2017, Nguyen2019}
enables a wide range of intriguing electrical phenomena including
Coulomb drag effects~\cite{Gorbachev2012, Seyoung2011, Song2012},
flat-band superconductivity~\cite{Wang2020, Cao2018},
and interlayer exciton superfluidity~\cite{Li2017, Liu2017, Wang2019}.

Among them, plasmons, which are self-sustained charge density oscillations, in 2D systems have been the focus of active research in recent years due to their influence on the optical and electronic properties of materials~\cite{Stern1967, Allen1977, Hwang2007, Fei2012, Grigorenko2012, Chen2012, Fei2015, Novelli2020}.
In the presence of a magnetic field, they are called magnetoplasmons which are generated by the collective excitations between Landau levels~\cite{Bernstein1958, Chiu1974, DasSarma1982, Kallin1984}. 
The splitting of plasmon modes due to the cyclotron motion of electrons has been envisioned for promising applications to nanophotonics~\cite{Bonanni2011, Melander2012, Pineider2013, Armelles2013, Han2017, Maccaferri2015}.
Furthermore, the possibility of plasmonic excitations to low frequencies has allowed extreme electromagnetic field confinement in multilayer nanostructures, resulting in ultra-strong light-matter coupling~\cite{Zhang2016, Keller2017}
and pronounced nonlocal effects~\cite{Grigelionis2015, Lundeberg2017, Cao2017}.
This calls for a systematic understanding of magnetoplasmons in multilayer structures.

\begin{figure}[htb!]
    \centering
    \includegraphics[width=1.0\linewidth]{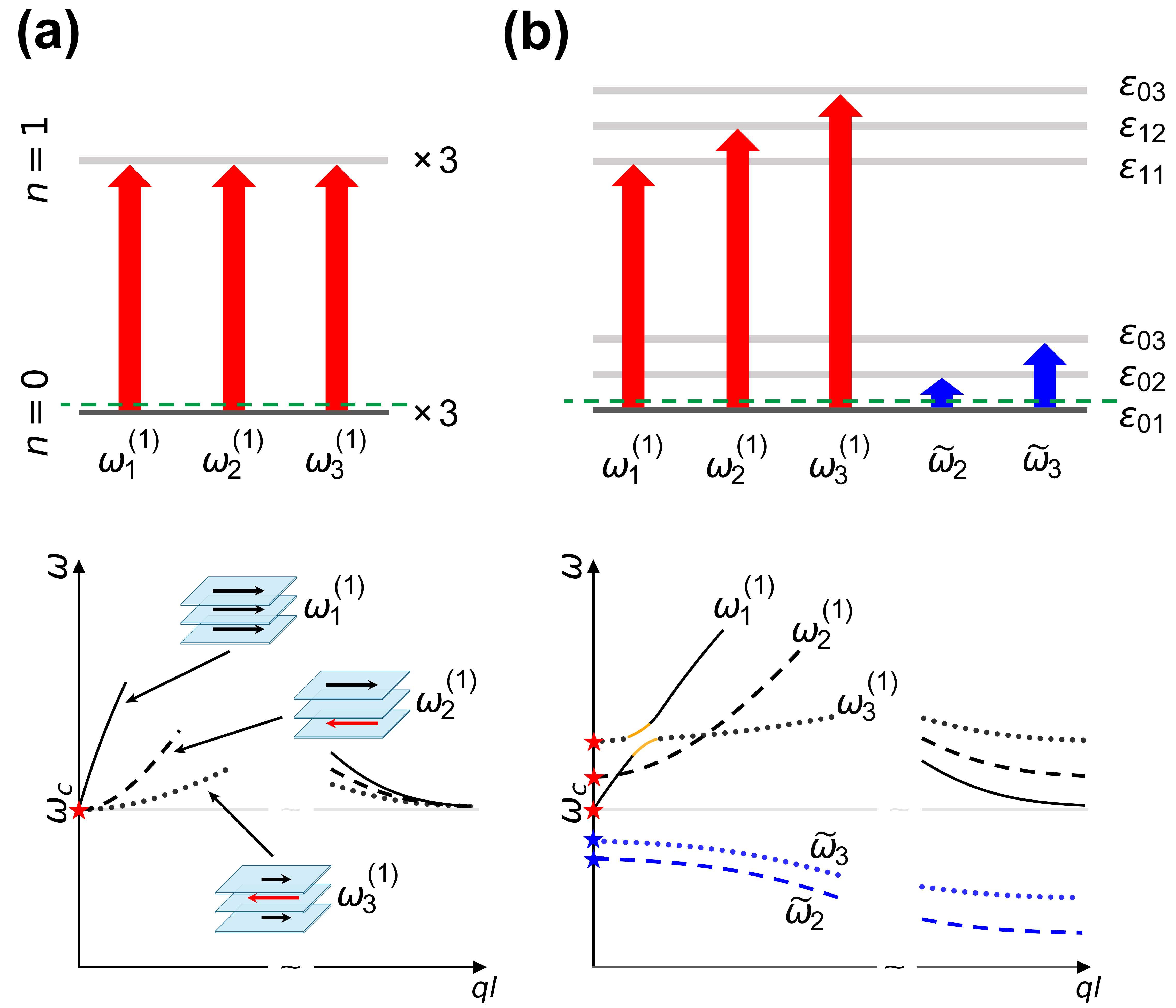}
    \caption{
    Schematic illustration of the dominant interband transitions for each mode and the corresponding mode dispersions in a trilayer system filled up to the lowest band: (a) the decoupled-layer limit and (b) the case with finite interlayer tunneling. The green dashed lines denote the Fermi energy, while the black (gray) horizontal lines represent occupied (unoccupied) Landau levels. The red (blue) arrows indicate the dominant transitions in the long-wavelength limit for the magnetoplasmon (tunneling magnetoplasmon) modes. The lower panels show the mode dispersions in the limits of $ql\to0$ and $ql\to\infty$. For a given mode, the same line style is used consistently across the figures. 
    The insets in (a) show the corresponding charge oscillations in the long-wavelength limit.
    In (b), magnetoplasmon (tunneling magnetoplasmon) modes are depicted by black (blue) lines, and anticrossing behavior between modes is highlighted in yellow.
    }
    \label{Fig: schm. pic for N=3}
\end{figure}

In this work, we theoretically investigate magnetoplasmon dispersions by incorporating the Kac--Murdock--Szeg\H{o} (KMS) Toeplitz matrix into the Landau levels and applying it to systems both with and without interlayer tunneling.
We find that, as in plasmons, the magnetoplasmon dispersion branch in the single-layer case splits into $N$ branches because long-range Coulomb coupling between the layers allows multiple independent collective charge oscillations: 
one mode in which the charge densities in all layers oscillate in phase, and $N-1$ modes in which the charge densities oscillate out of phase.
We derive the asymptotic behavior of each dispersion in the long-wavelength limit, as well as in the limit of large layer separation and strong magnetic fields.
When interlayer tunneling is present, the Landau levels split, and the resulting interband transitions qualitatively modify the magnetoplasmon spectrum.
We provide a systematic framework to clarify the spectrum, including the number of modes and their gaps, by identifying the magnetoplasmon mode associated with each interband transition and the hybridization between these modes governed by underlying symmetries.
Moreover, interband transitions with the same Landau level index give rise to tunneling magnetoplasmons~\cite{Aizin1995}, which acquire an enhanced plasmon gap exceeding the associated interband gap due to the Coulomb interaction.
We obtain this correction analytically in the weak Coulomb interaction limit, where the interaction is negligibly small compared to the kinetic energy.
Our results for systems filled up to the lowest bands are schematically illustrated in Fig.~\ref{Fig: schm. pic for N=3} for the trilayer case.

The paper is organized as follows.
In Sec.~\ref{Sec:RPA_KMS}, we present the basic formulation of our approach, including the random phase approximation (RPA) and the KMS structure of the Coulomb interaction in the system. The selection rules arising from the system symmetry and interaction structure are also discussed.
Section~\ref{Sec: Results} is devoted to our results for two-dimensional electron gas (2DEG) systems, covering both decoupled cases (Sec.~\ref{Sec: Results}.A) and coupled cases (Sec.~\ref{Sec: Results}.B), with particular emphasis on trilayer systems.
Finally, the extension to multilayer 2DEG systems with $N>3$ and to multilayer graphene is discussed in Sec.~\ref{Sec: Discussion}.

\section{Theoretical Background} \label{Sec:RPA_KMS}
\begin{table*}[t]
\caption{Asymptotic forms of the eigenvalues $g_\alpha(e^{-qd})$ and the eigenvectors
$\mathbf{u}_\alpha(e^{-qd})=\bigl(u_\alpha^{(1)},\dots,u_\alpha^{(N)}\bigr)^{\mathsf T}$ of the KMS matrix $A_{ij}(e^{-qd})$.
Here $\alpha=1,\dots,N$ and $k=1,\dots,N$.
The eigenvectors $\mathbf{u}_\alpha$ are presented up to a normalization factor.}
\label{Tab:KMS_asymptotics}

\renewcommand{\arraystretch}{1.5}

\newcommand{\Cen}[1]{\parbox[c]{0.42\textwidth}{\centering $\displaystyle #1$}}

\begin{ruledtabular}
\begin{tabular*}{\textwidth}{@{\extracolsep{\fill}} l c c }
Limit & \Cen{g_\alpha(e^{-qd})} & \Cen{u_\alpha^{(j)}(e^{-qd})} \\
\hline
${qd\to 0}$ &
\Cen{g_{1}=N,\quad g_{\alpha}=\frac{qd}{1-\cos\!\bigl(\frac{\alpha-1}{N}\pi\bigr)}\ (2\le\alpha\le N)} &
\Cen{\cos\!\Bigl[\frac{(2j-1)(\alpha-1)}{2N}\pi\Bigr]} \\
${qd\to \infty}$ &
\Cen{g_{\alpha} \simeq 1+2e^{-qd}\cos\!\Bigl(\frac{\alpha\pi}{N+1}\Bigr)} &
\Cen{\sin\!\Bigl(\frac{\alpha j\pi}{N+1}\Bigr)}
\end{tabular*}
\end{ruledtabular}
\end{table*}

In this paper, we consider an $N$-layer system with interlayer separation $d$ under a magnetic field $B$ applied perpendicular to the layers.
We retain only nearest-neighbor interlayer tunneling with amplitude $t$.
An eigenstate $\lvert n,\lambda \rangle$ of the system, labeled by a Landau level index $n$ and a subband index $\lambda=1,\dots,N$, can be written as
\begin{equation} \label{Eq:subband wavefunction}
\lvert n,\lambda\rangle
= \sqrt{\frac{2}{N+1}}
\sum_{j=1}^{N} \psi^{(j)}_\lambda\,
\lvert n\rangle_{j},
\end{equation}
where $\psi^{(j)}_\lambda = (-1)^j
\sin\!\left(\frac{j\lambda\pi}{N+1}\right)$ is the subband wave function,
and $\lvert n\rangle_j$ denotes the Landau level-$n$ state on the $j$th layer.
The corresponding eigenenergy is
\( \varepsilon_{n\lambda}
= \left(n+\frac{1}{2}\right)\hbar\omega_c + \Delta_\lambda \)
with $\Delta_\lambda=-2t\cos\left(\frac{\lambda\pi}{N+1}\right)$.

The noninteracting density--density response function of the system at temperature $T$ reads~\cite{mahan1990many,Giuliani_Vignale_2005}
\begin{eqnarray}
    \chi_{ij}^{(0)}(\mathbf{q}, \omega)
    &=& \frac{g_s}{2 \pi l^2}
    \sum_{n,n',\lambda,\lambda'}
    \frac{f(\varepsilon_{n\lambda})-f(\varepsilon_{n'\lambda'}) }
    {\hbar\omega+\varepsilon_{n\lambda}-\varepsilon_{n'\lambda'}+i0^+} \nonumber\\
    &\times& F_{nn'}(\mathbf{q})P^{\lambda\lambda'}_{ij}
\label{Eq: χ0_layerbasis}
\end{eqnarray}
in the layer basis.
Here, $g_s=2$ is the spin degeneracy, $l=\sqrt{\hbar c/(eB)}$ is the magnetic length,
$\omega_c=eB/(mc)$ is the cyclotron frequency, $\varepsilon_F$ is the Fermi energy,
and $f(\varepsilon)= \left[e^{(\varepsilon-\varepsilon_F)/(k_BT)}+1\right]^{-1}$ is the Fermi-Dirac distribution function.
The factor
\begin{equation}
P^{\lambda\lambda'}_{ij}
=\langle \lambda | P_i | \lambda' \rangle
 \langle \lambda' | P_j | \lambda \rangle
 \label{Eq: overlap_layer}
\end{equation}
is the wave function overlap factor, where $P_i$ denotes the projector onto the $i$th layer.
The form factor $F_{nn'}(\mathbf{q})$ of a 2DEG is given by~\cite{Giuliani_Vignale_2005,Roldan2009}
\begin{equation}
F_{n,n'}(\mathbf{q})
= e^{-q^2l^2/2}
\left(\frac{q^2l^2}{2}\right)^{|n-n'|}
\frac{n_<!}{n_>!}
\left[
L_{n_<}^{\,|n-n'|}
\!\left(\frac{q^2l^2}{2}\right)
\right]^2,
\label{Eq: 2DEG form factor}
\end{equation}
with $n_>=\max(n,n')$, $n_<=\min(n,n')$, and $L_a^{\,b}(x)$ denoting the associated Laguerre polynomials.

$N$-layer magnetoplasmon dispersions are determined by the condition
$\det[\epsilon_{ij}(\mathbf{q},\omega)]=0$, where $\epsilon_{ij}$ is the dielectric matrix, given within the RPA by
\begin{equation}
\epsilon_{ij}(\mathbf{q},\omega)
= \delta_{ij}-\sum_{k}V_{ik}(q)\,\chi^{(0)}_{kj}(\mathbf{q},\omega).
\label{Eq:epsilon_layer_basis}
\end{equation}
$V_{ij}(q)=v(q)\,e^{-|i-j|qd}$ denotes the Coulomb interaction between the $i$th layer and $j$th layer, with
$v(q)=2\pi e^{2}/(\kappa q)$ and $\kappa$ the background dielectric constant.
For each dispersion $\omega(\mathbf{q})$, the eigenvector of the dielectric matrix with a vanishing eigenvalue describes how the oscillation is distributed among the layers through its coefficients in the layer basis.

The Coulomb matrix $V_{ij}(q)$ can be written in terms of the KMS matrix $A_{ij}(\rho)=\rho^{|i-j|}$ ($0<\rho<1$) as $V_{ij}(q)=v(q)A_{ij}(\rho)$ with $\rho=e^{-qd}$, which allows for a tractable analysis of the plasmon dispersions in an $N$-layer system.
The eigenvalues $g_\alpha(e^{-qd})$ and eigenvectors $\mathbf{u}_\alpha(e^{-qd})=\bigl(u_\alpha^{(1)},\dots,u_\alpha^{(N)}\bigr)^{\mathsf T}$ ($\alpha=1,\dots,N$) of $A_{ij}(e^{-qd})$ exhibit well-defined asymptotic forms in both the $qd\to0$ and $qd\to\infty$ limits, as summarized in Table~\ref{Tab:KMS_asymptotics}.
For example, for $N=3$ and $\alpha=3$ in the $qd\to0$ limit, $\mathbf{u}_{\alpha=3}(e^{-qd}\rightarrow 1)=\frac{1}{\sqrt{6}}(1,-2,1)^{\mathsf T}$, indicating that the first and third layers oscillate in the same direction while the second layer oscillates in the opposite direction with twice their amplitude (see App.~\ref{App.KMS} for the explicit expressions of the eigenpairs).
In the basis of Coulomb eigenvectors, which diagonalize $V_{ij}$, the dielectric matrix takes the form
\begin{equation}
\epsilon_{\alpha\beta}(\mathbf{q},\omega)
= \delta_{\alpha\beta} - V_\alpha(q)\,\chi^{(0)}_{\alpha\beta}(\mathbf{q},\omega),
\label{Eq:epsilon_Coulomb_basis}
\end{equation}
where $V_\alpha(q)=v(q)\,g_\alpha(e^{-qd})$.
The susceptibility $\chi^{(0)}_{\alpha\beta}$ is obtained from Eq.~(\ref{Eq: χ0_layerbasis}) by replacing
$P^{\lambda\lambda'}_{ij}$ in Eq.~(\ref{Eq: overlap_layer}) with
\begin{equation}
P^{\lambda\lambda'}_{\alpha\beta}
= \langle \lambda | U_\alpha | \lambda' \rangle
  \langle \lambda' | U_\beta | \lambda \rangle,
\label{Eq: overlap_Coulomb}    
\end{equation}
where $U_\alpha=\mathrm{diag}\!\bigl[\mathbf{u}_\alpha(e^{-qd})\bigr]$.

Coulomb eigenvectors $\mathbf{u}_\alpha$ exhibit symmetry under layer inversion, which provides a selection rule for interband-transition contributions to Coulomb oscillations~\cite{wvhd-492f}.
Odd-index Coulomb eigenvectors are symmetric under layer inversion about the midplane, whereas even-index eigenvectors are antisymmetric.
In a system whose bands are either symmetric or antisymmetric, interband transitions between a symmetric and an antisymmetric band contribute exclusively to antisymmetric Coulomb oscillations, whereas all other transitions contribute to symmetric Coulomb oscillations.
In our system, the odd (even) subband index corresponds to a symmetric (antisymmetric) band with respect to layer inversion, and thus band transitions between subband indices of the same (opposite) parity contribute only to Coulomb oscillations of odd (even) index.

In addition to the symmetry constraint, a further selection rule is present in our system.
The term $\langle \lambda | U_\alpha | \lambda' \rangle$ is negligible in the long-wavelength limit unless
$|\lambda-\lambda'|=\alpha-1$ or $|\lambda+\lambda'-(N+1)|=N+1-(\alpha-1)$. 
For example, the matrix element $\langle \lambda=1 \lvert U_{\beta} \rvert \lambda'=\alpha \rangle$ is appreciable only for $\beta=\alpha$ and $\beta=\alpha+2$; both are allowed by the selection rule, but the former is numerically larger in practice.
See Sec.~\RNum{3} of the Supplemental Material (SM) in Ref.~\cite{wvhd-492f} for the explicit derivation.
The quantity $P_{\alpha\beta}^{\lambda\lambda'}$ is dominant when both $\alpha$ and $\beta$ satisfy this condition for given $\lambda$ and $\lambda'$.
This rule specifies the Coulomb eigenmodes to which a given interband transition predominantly contributes. 

\section{Results} \label{Sec: Results}
\subsection{Decoupled system}
When there is no interlayer tunneling and all $N$ layers are filled up to the same filling factor $\nu$, the noninteracting density response function is diagonal in the layer basis,
$\chi_{ij}^{(0)}=\delta_{ij}\chi_{0}$,
where $\chi_{0}$ denotes the noninteracting response function of a single-layer 2DEG.
As a result, the Coulomb eigenvectors diagonalize $\epsilon_{\alpha\beta}$, and the magnetoplasmon dispersion branch that emanates from each $k\omega_c$ ($k=1,2,\dots$) in the single-layer case splits into $N$ branches, denoted by $\omega_\alpha^{(k)}$ ($\alpha=1,\dots,N$) with the associated Coulomb eigenvector $\mathbf{u}_\alpha$.

In the long-wavelength limit, the dispersion $\omega_\alpha^{(k)}$ is governed solely by interband transitions with gap $k\hbar\omega_c$.
Incorporating only these interband transitions, closed-form expressions for the dispersion $\omega_\alpha^{(k)}$ can be obtained as (see App.~\ref{App.derivation in decoupled limit} for the derivation)

\begin{align}
[\omega_\alpha^{(k)}(q )]^2
&\xrightarrow[ qd,ql\to 0]{}
k^2\omega_c^2
\label{Eq:small_q_chg}  \\[-2pt]
&+ \frac{k A_k g_s e^2}{\kappa m}\,
\begin{cases}
\frac{\sqrt 2N}{l^3}\,\left(\frac{q^2l^2}{2}\right)^{k-1/2}, & \alpha=1,\\[4pt]
\frac{2d/l^4}{1-\cos\!\bigl[\pi(\alpha-1)/N\bigr]}\,\left(\frac{q^2l^2}{2}\right)^{k}, & \alpha\neq 1,
\end{cases} \nonumber
\end{align}
and in the limit of large layer separation and strong magnetic fields, i.e., $qd, \, ql \to \infty$, the magnetoplasmons behave like
\begin{align}
[\omega_\alpha^{(k)}&(q)]^2
\xrightarrow[qd,ql\to \infty]{}
k^2\omega_c^2 +\left[1+2e^{-qd}\cos\left(\frac{\alpha \pi}{N+1}\right)\right] \notag\\&\times
\frac{\,e^{-q^2l^2/2}}{(\nu+n-1)!(\nu-1)!}\frac{k g_s e^2}{\kappa m }\frac{\sqrt{2}}{l^3} \left(\frac{q^2l^2}{2}\right)^{k+2\nu-5/2},
\label{Eq:large_q_chg}
\end{align}
where $A_k \equiv \sum_{j =j_{k}}^{\nu-1} \frac{(j+k)!}{j!(k!)^2}$ and $j_{k}=\max(0,\nu-k)$.
Note that the $k=1$ branches carry the lowest power of $q$ and therefore govern the leading long-wavelength behavior in Eq.~(\ref{Eq:small_q_chg}).
The in-phase mode ($\alpha=1$) within these branches shows a linear dependence on $q$ in the long-wavelength limit, rather than the conventional $\sqrt{q}$ behavior of the non-magnetic mode.

\begin{figure}
    \centering
    \includegraphics[width=1.0\linewidth]{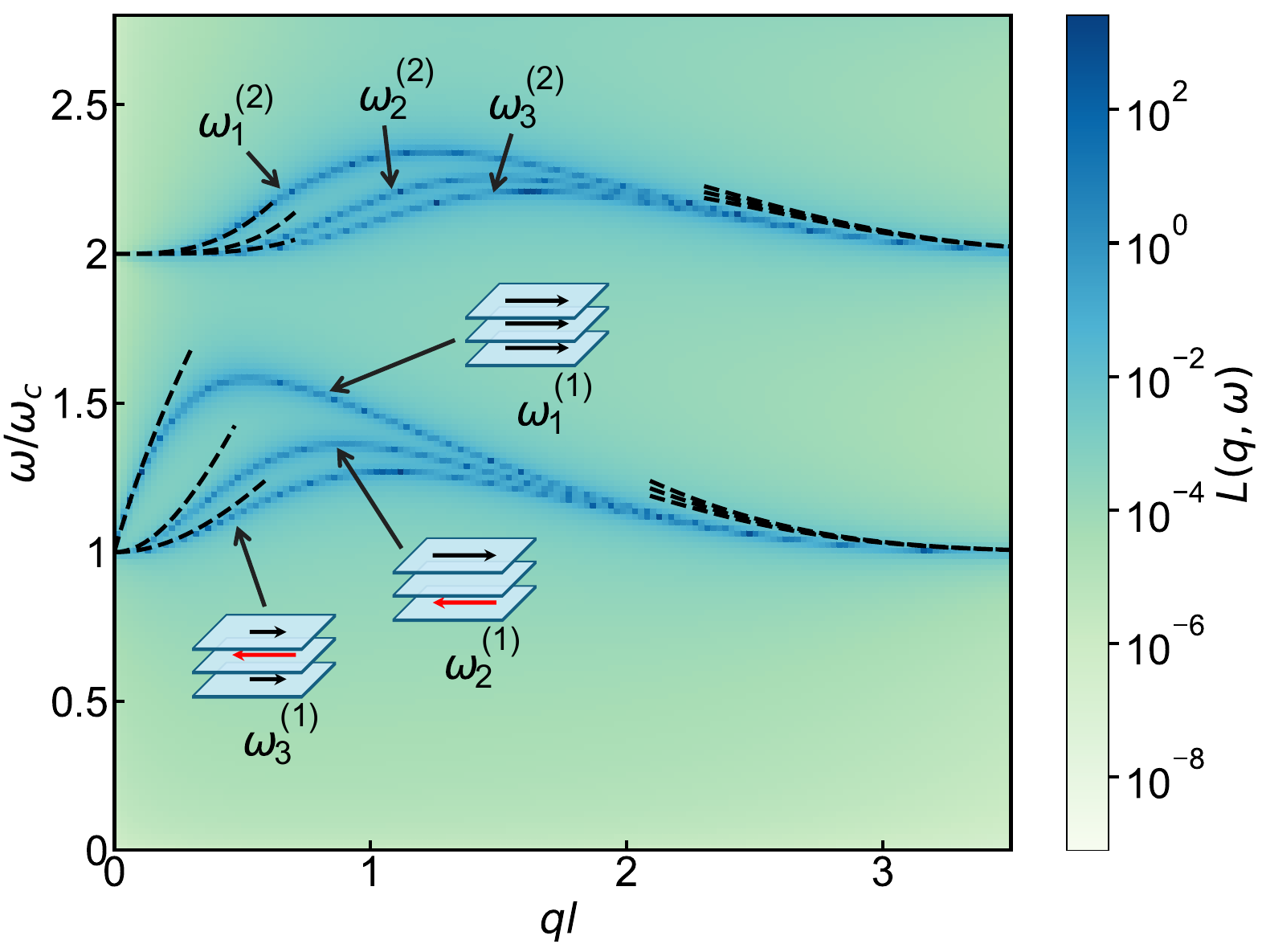} 
    \caption{Loss function plot $L(\mathbf{q},\omega)=-\text{Im}{\text{Tr}[\epsilon^{-1}(\mathbf{q},\omega)]}$ for a trilayer system in the decoupled limit, compared with the asymptotic forms in Eq.~(\ref{Eq:small_q_chg}) and Eq.~(\ref{Eq:large_q_chg}), which are drawn as black dashed lines.
    The inset displays Coulomb oscillations for the $\omega_\alpha^{(1)}$ ($\alpha=1,2,3$) modes; note that $\omega_\alpha^{(2)}$ exhibits an identical oscillation pattern.
    For the calculations, we used parameters typical for a GaAs quantum well: $m=0.067\,m_e$, $\kappa=10.9$, $d=100\,\text{\AA}$, and $\hbar\omega_c=15\,\text{meV}$, with a phenomenological broadening $\eta=10^{-3}\,\hbar\omega_c$.}
    \label{Fig: decoupled N=3}
\end{figure}

We present numerical results for a decoupled trilayer system in Fig.~\ref{Fig: decoupled N=3} as an example.
The asymptotic forms of $\omega_\alpha^{(k)}$ given by Eq.~(\ref{Eq:small_q_chg}) and Eq.~(\ref{Eq:large_q_chg}) agree with the numerical results.
In general, for $k=1$ the agreement between the numerical results and Eq.~(\ref{Eq:small_q_chg}) is robust regardless of the magnetic-field strength, whereas for $k\neq1$ the agreement becomes more evident as the magnetic field increases.
The reason is that, for each $\omega_\alpha^{(k\neq1)}$, the range of $q$ over which the transitions with band gap $k\hbar\omega_c$ provide the dominant contribution to the dispersion shrinks as the magnetic field increases.
Typically, the parameter $\frac{e^2}{\kappa l}/(\hbar\omega_c)$ controls the range of validity of the asymptotic behavior; thus, the agreement improves at higher magnetic fields or smaller $\kappa$.

\subsection{Coupled system}
Unlike the decoupled case, interlayer tunneling $t$ lifts the Landau level degeneracy by splitting each level into multiple subbands.
As a result, the Fermi energy $\varepsilon_F$ can lie between the bands sharing the same Landau level index $n$, resulting in a partially filled Landau level and enabling intersubband transitions within the same Landau level to become allowed.
To simplify the band sequence in energy, we primarily focus on the limit $t \ll \hbar\omega_{c}$, as illustrated in Fig.~\ref{Fig: schm. pic for N=3}.
Note that the qualitative features of our results persist for $t \gtrsim \hbar\omega_c$, as discussed in the Sec.~\ref{Sec: Discussion}.
\begin{figure*}[t]
    \centering
    \includegraphics[width=1.0\linewidth]{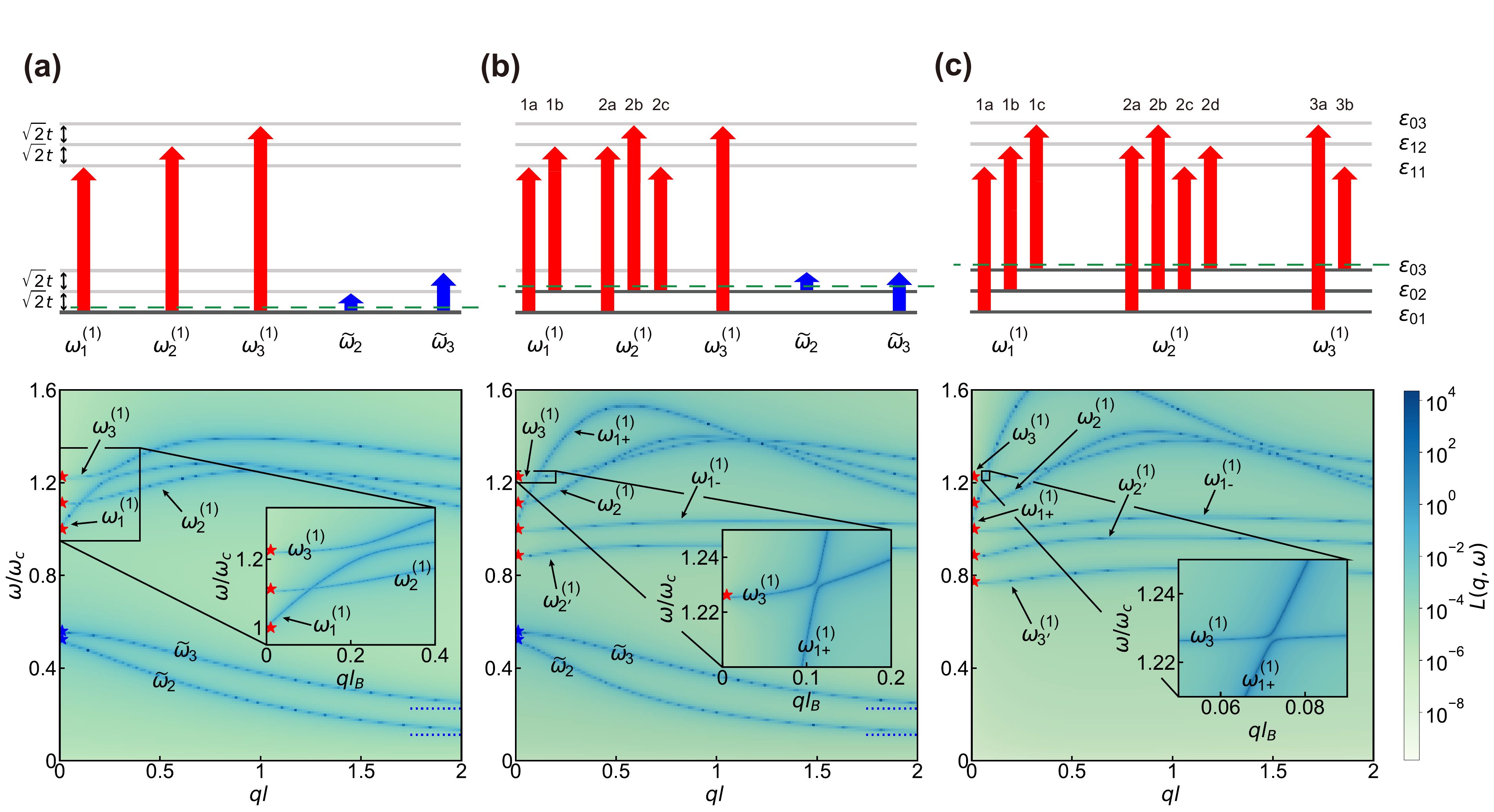}
    \caption{
Schematic illustration of magnetoplasmons associated with each band transition and the loss function plot $L(\mathbf{q},\omega)=-\text{Im}{\text{Tr}[\epsilon^{-1}(\mathbf{q},\omega)]}$ for a coupled trilayer system filled up to the states
(a) $\varepsilon_{01}$, (b) $\varepsilon_{02}$, and (c) $\varepsilon_{03}$.
For band transitions indicated by red (blue) arrows, the corresponding magnetoplasmons (tunneling magnetoplasmons) are listed above the arrows.
When multiple magnetoplasmons share the same gap, a splitting occurs unless the modes are related by subband-index-reversal symmetry.
For example, in (b), $\omega_{1a}^{(1)}$ and $\omega_{1b}^{(1)}$ combine to form $\omega_{1,\pm}^{(1)}$, whereas $\omega_{2a}^{(1)}$ and $\omega_{2b}^{(1)}$ are degenerate and form only $\omega_{2}^{(1)}$.
The red and blue stars indicate the magnetoplasmon and tunneling magnetoplasmon gap values evaluated using Eq.~(\ref{Eq:MP_gap}) for the red stars in (a), Eq.~(\ref{Eq:MP_gap_modified}) for the red stars in (b) and (c), and Eq.~(\ref{Eq:TMP_mode_gap}) for the blue stars.
The blue dashed lines represent Eq.~(\ref{Eq:TMP_mode_infty}).
The insets in each panel highlight the crossing and anticrossing behaviors between modes; the corresponding regions are indicated by black boxes.
The calculations were performed using the same parameters as in Fig.~(\ref{Fig: schm. pic for N=3}), except that $t=0.8\,\mathrm{meV}$.
}
    \label{Fig:numerical_results_N3_coupled}
\end{figure*}

We first focus on the cases where the system is filled up to the band $\varepsilon_{01}$, so that only the lowest subband $\lambda=1$ of the $n=0$ Landau level is fully occupied.
The interband transitions between bands with different Landau level indices result in magnetoplasmon dispersions similar to those in the decoupled case.
They can be labeled by their gap values as
\begin{equation}
  \hbar\omega_{\alpha}^{(k)}(ql \to 0) = \Delta_{\alpha 1}+ k\hbar \omega_c,
  \label{Eq:MP_gap}
\end{equation}
with a positive integer $k,\,\alpha=1,\dots,N$, and $\Delta_{ij}=\Delta_i-\Delta_j$.
The corresponding dominant interband transition in the long-wavelength limit occurs between the band $\varepsilon_{01}$ and the band $\varepsilon_{k\alpha}$.
As the inversion symmetry is determined by the parity of $\alpha$, the mode $\omega_\alpha^{(k)}$ follows the parity of $\alpha$.
Thus, we can expect crossing (anticrossing) behavior between modes $\omega_\alpha^{(k)}$ with opposite (same) $\alpha$ parity.
Moreover, the transition between $\varepsilon_{01}$ and $\varepsilon_{k\alpha}$ predominantly contributes to the Coulomb oscillations associated with $\mathbf{u}_\alpha$ as dictated by the selection rule. 

As a specific example, we consider a coupled trilayer system filled up to the state $\varepsilon_{01}$ [see Fig.~\ref{Fig:numerical_results_N3_coupled}(a)].
Three dispersions at and above $\omega_c$ correspond to the modes
$\omega_\alpha^{(1)}$ ($\alpha=1,2,3$) and show exact agreement with Eq.~(\ref{Eq:MP_gap}).
Moreover, anticrossing behavior is observed between $\omega_1^{(1)}$ and $\omega_3^{(1)}$, while they both cross with $\omega_2^{(1)}$.

The remaining interband transitions between the bands with the same Landau level index give rise to the tunneling magnetoplasmon modes, $\tilde{\omega}_\alpha$ ($\alpha=2,\dots,N$).
Tunneling magnetoplasmon modes acquire an additional gap enhancement relative to the underlying interband gap due to the Coulomb interaction unlike other magnetoplasmon modes $\omega_{\alpha}^{(k)}$.
This property can be checked analytically in the weak Coulomb interaction limit, defined by the condition $(e^2/\kappa l)/t \ll 1$.
In this limit, the dominant interband transition contributing to $\tilde{\omega}_{\alpha}$ occurs between $\varepsilon_{01}$ and $\varepsilon_{0\alpha}$ in both the $ql\to0$ and $ql\to\infty$ limits.
The values of $\tilde{\omega}_\alpha$ in these limits are given by (see App.~C for the explicit derivation)
\begin{subequations} \label{Eq:TMP}
\begin{align}
    [\hbar\tilde{\omega}_{\alpha}(ql\to 0)]^2
    &= \Delta_{\alpha1}^2
    + \bigl(V_\alpha P_{\alpha,\alpha}^{\alpha 1} + V_{\alpha+2} P_{\alpha+2,\alpha+2}^{\alpha 1}\bigr)
    \frac{g_s \Delta_{\alpha1}}{\pi l^2},
    \label{Eq:TMP_mode_gap}\\
    \hbar\tilde{\omega}_{\alpha}(ql\to \infty) &= \Delta_{\alpha1},
    \label{Eq:TMP_mode_infty}
\end{align}
\end{subequations}
where we set $V_\alpha=0$ and $P_{\alpha,\alpha}^{\alpha 1}=0$ for $\alpha>N$.
Here, $V_\alpha(qd\to0)$ is finite because $g_\alpha(e^{-qd})$ has a linear dependence on $qd$ when $\alpha\neq1$.

It is worth emphasizing that Eq.~(\ref{Eq:TMP}) remains valid for $N=3$ even when the weak Coulomb interaction limit is not satisfied. In Fig.~\ref{Fig:numerical_results_N3_coupled}(a), the two dispersions below $\omega_{c}$ correspond to tunneling magnetoplasmons $\tilde{\omega}_{\alpha}$. Both are well described by Eq.~(\ref{Eq:TMP}), even though the Coulomb interaction is not weak.
This agreement originates from the selection rule in the Coulomb eigenbasis, which ensures that each dispersion $\tilde{\omega}_{\alpha}$ arises from the interband transition between $\varepsilon_{01}$ and $\varepsilon_{0\alpha}$.
Specifically, the transition between $\varepsilon_{01}$ and $\varepsilon_{03}$ ($\tilde{\omega}_3$) involves symmetric bands and therefore couples only to odd Coulomb oscillations, whereas the transition between $\varepsilon_{01}$ and $\varepsilon_{02}$ ($\tilde{\omega}_2$) couples only to an even Coulomb oscillation.

We now consider the case in which higher bands are occupied. When multiple bands are filled, more than one interband transition corresponds to the same plasmon gap. As a result, several magnetoplasmon branches emerge from the same gap, originating from different transitions and subsequently hybridizing and splitting into distinct collective modes.
Accordingly, Eq.~(\ref{Eq:MP_gap}) can be generalized to
\begin{equation}
\hbar\omega_\alpha^{(k)}(ql\to 0) = \Delta_{\lambda'\lambda} + k\hbar\omega_c,
\label{Eq:MP_gap_modified}
\end{equation}
where $\lambda$ and $\lambda'$ denote the subband indices of the interband transitions between $\varepsilon_{n\lambda}$ and $\varepsilon_{n'=n+k,\lambda'}$ for a given band filling.

However, the number of such modes is generally nontrivial, since distinct interband transitions may support identical Coulomb oscillations.
This occurs when two transitions share the same Landau level indices and their subband indices are related by subband-index-reversal symmetry about the central subband,
defined by $\lambda \rightarrow N+1-\lambda$ and $\lambda' \rightarrow N+1-\lambda'$.
Specifically, these correspond to the transition between $\varepsilon_{n\lambda}$ and $\varepsilon_{n'\lambda'}$, and that between
$\varepsilon_{n,N+1-\lambda}$ and $\varepsilon_{n',N+1-\lambda'}$.
These two transitions share the same energy gap due to the relation $\Delta_\lambda = -\Delta_{N+1-\lambda}$, and the fact that they support identical Coulomb oscillations arises from the symmetry properties of
$P_{\alpha\beta}^{\lambda\lambda'}$ in Eq.~(\ref{Eq: overlap_layer}), which obeys
\begin{equation}
P_{\alpha\beta}^{\lambda\lambda'}
= P_{\alpha\beta}^{(N+1-\lambda)(N+1-\lambda')}
= P_{\alpha\beta}^{(N+1-\lambda')(N+1-\lambda)}.
\end{equation}
The first equality follows from the relation between the subband wave functions,
$\psi^{(j)}_\lambda = (-1)^{j+1}\psi^{(j)}_{N+1-\lambda}$,
while the second equality reflects the Hermiticity of
$P_{\alpha\beta}^{\lambda\lambda'}$, which is real and satisfies
$P_{\alpha\beta}^{\lambda\lambda'} = P_{\alpha\beta}^{\lambda'\lambda}$.
Since $P_{\alpha\beta}^{\lambda\lambda'}$ determines the dielectric matrix
$\epsilon_{\alpha\beta}$, distinct interband transitions sharing the same energy gap and identical
$P_{\alpha\beta}^{\lambda\lambda'}$ support the same Coulomb oscillations.

As an example, we first consider the case where a trilayer is filled up to the state $\varepsilon_{02}$ [see Fig.~\ref{Fig:numerical_results_N3_coupled}(b)].
In this example, we focus on the case $k=1$ in Eq.~(\ref{Eq:MP_gap_modified}).
The six interband transitions between $\varepsilon_{0\lambda}$ and $\varepsilon_{1\lambda'}$ can be grouped according to $|\lambda-\lambda'|=0,1,2$, and the corresponding magnetoplasmon modes are labeled as illustrated in the figure.
For example, $\omega_{1a}^{(1)}$ corresponds to the magnetoplasmon arising from the transition between $\varepsilon_{01}$ and $\varepsilon_{11}$, whereas $\omega_{1b}^{(1)}$ corresponds to the transition between $\varepsilon_{02}$ and $\varepsilon_{12}$. 

The modes $\omega_{1a}^{(1)}$ and $\omega_{1b}^{(1)}$ share the same gap value $\hbar\omega_{c}$ and hybridize to form the magnetoplasmon dispersions $\omega_{1\pm}^{(1)}$, which emerge from $\omega_{c}$.
In contrast, the band transitions associated with the magnetoplasmons $\omega_{2a}^{(1)}$ and $\omega_{2b}^{(1)}$ are related by subband-index-reversal symmetry; therefore, no splitting occurs and a unique mode $\omega_{2}^{(1)}$ emerges from $\omega_{c}+\sqrt{2}t/\hbar$.
The mode $\omega_{2c}^{(1)}$ does not have a transition with the same gap and thus remains as a single branch $\omega_{2'}^{(1)}$ emerging from $\omega_{c}-\sqrt{2}t/\hbar$, similar to $\omega_{3}^{(1)}$, which emerges from $\omega_{c}+2\sqrt{2}t/\hbar$.
As a result, the six transitions give rise to a total of five modes, all of which exhibit gaps consistent with Eq.~(\ref{Eq:MP_gap_modified}).
As in the case where only the lowest band is filled, anticrossing behavior is observed between modes with the same parity, such as $\omega_{1+}^{(1)}$ and $\omega_{3}^{(1)}$.

Since interband transitions with the same Landau level index $n$ are present in this example, tunneling magnetoplasmons appear.
The symmetry rule indicates that the tunneling magnetoplasmons in this system are identical, in the long-wavelength limit, to those in the case where only the lowest band is filled.
This equivalence follows from the fact that the band transition between $\varepsilon_{02}$ and $\varepsilon_{03}$ is mapped onto the transition between $\varepsilon_{01}$ and $\varepsilon_{02}$ by subband-index-reversal symmetry.
As a result, Eq.~(\ref{Eq:TMP}) can be used to describe the tunneling magnetoplasmons shown in Fig.~\ref{Fig:numerical_results_N3_coupled}(b).

Next, we consider a trilayer filled up to $\varepsilon_{03}$ [see Fig.~\ref{Fig:numerical_results_N3_coupled}(c)]. Nine interband transitions between $\varepsilon_{0\lambda}$ and $\varepsilon_{1\lambda'}$ are available in the system.
Grouping the modes as in the previous example, $\omega_{1a}^{(1)}$, $\omega_{1b}^{(1)}$, and $\omega_{1c}^{(1)}$ are the three magnetoplasmons originating from the band transitions across the gap $\hbar\omega_c$.
Owing to subband-index-reversal symmetry, $\omega_{1a}^{(1)}$ and $\omega_{1c}^{(1)}$ are identical; these modes hybridize with $\omega_{1b}^{(1)}$, giving rise to $\omega_{1\pm}^{(1)}$.
Moreover, $\omega_{2a}^{(1)}$ and $\omega_{2b}^{(1)}$ are identical and together form $\omega_{2}^{(1)}$, while $\omega_{2c}^{(1)}$ and $\omega_{2d}^{(1)}$ are identical, forming $\omega_{2'}^{(1)}$; both cases come from subband-index-reversal symmetry.
In contrast, $\omega_{3a}^{(1)}$ and $\omega_{3b}^{(1)}$ have no transitions with the same gap and therefore give rise to $\omega_{3}^{(1)}$ and $\omega_{3'}^{(1)}$, respectively.
In total, six magnetoplasmons emerge from the nine transitions. All gap positions coincide with those in the previous example except for $\omega_{3'}^{(1)}$, which emerges from $\omega_c - 2\sqrt{2}t/\hbar$, in agreement with Eq.~(\ref{Eq:MP_gap_modified}).
No tunneling magnetoplasmon modes exist in this case, since there is no available band transition between $\varepsilon_{0\lambda}$ and $\varepsilon_{0\lambda'}$.
The general ideas illustrated by these examples can be straightforwardly extended to multilayer systems, which will be discussed in the next section.

\section{Discussion and conclusion} \label{Sec: Discussion}
In this section, we discuss the magnetoplasmon dispersions in $N$-layer 2DEG systems with $N>3$, as well as those in graphene multilayer systems.
In both cases, we observe crossing (anticrossing) behaviors between magnetoplasmon modes with opposite (same) parity, together with the emergence of tunneling magnetoplasmons.
Our key results for identifying individual modes in coupled $N$-layer systems, based on the hybridization between magnetoplasmon modes associated with different band transitions, extend directly to these systems.

We first consider $N$-layer 2DEG systems with $N>3$ in the presence of interlayer tunneling.
(The description of the decoupled case for general $N$ is given in Sec.~\ref{Sec: Results}.A.)
When the system is filled only up to the lowest band, no qualitative differences arise in the magnetoplasmon modes $\omega_{\alpha}^{(k)}$ described by Eq.~(\ref{Eq:MP_gap}).
By contrast, additional corrections are generally required for tunneling magnetoplasmons in Eq.~(\ref{Eq:TMP_mode_gap}),
since the assumption of a single dominant interband transition is no longer valid.

\begin{figure}
    \centering
    \includegraphics[width=1.0\linewidth]{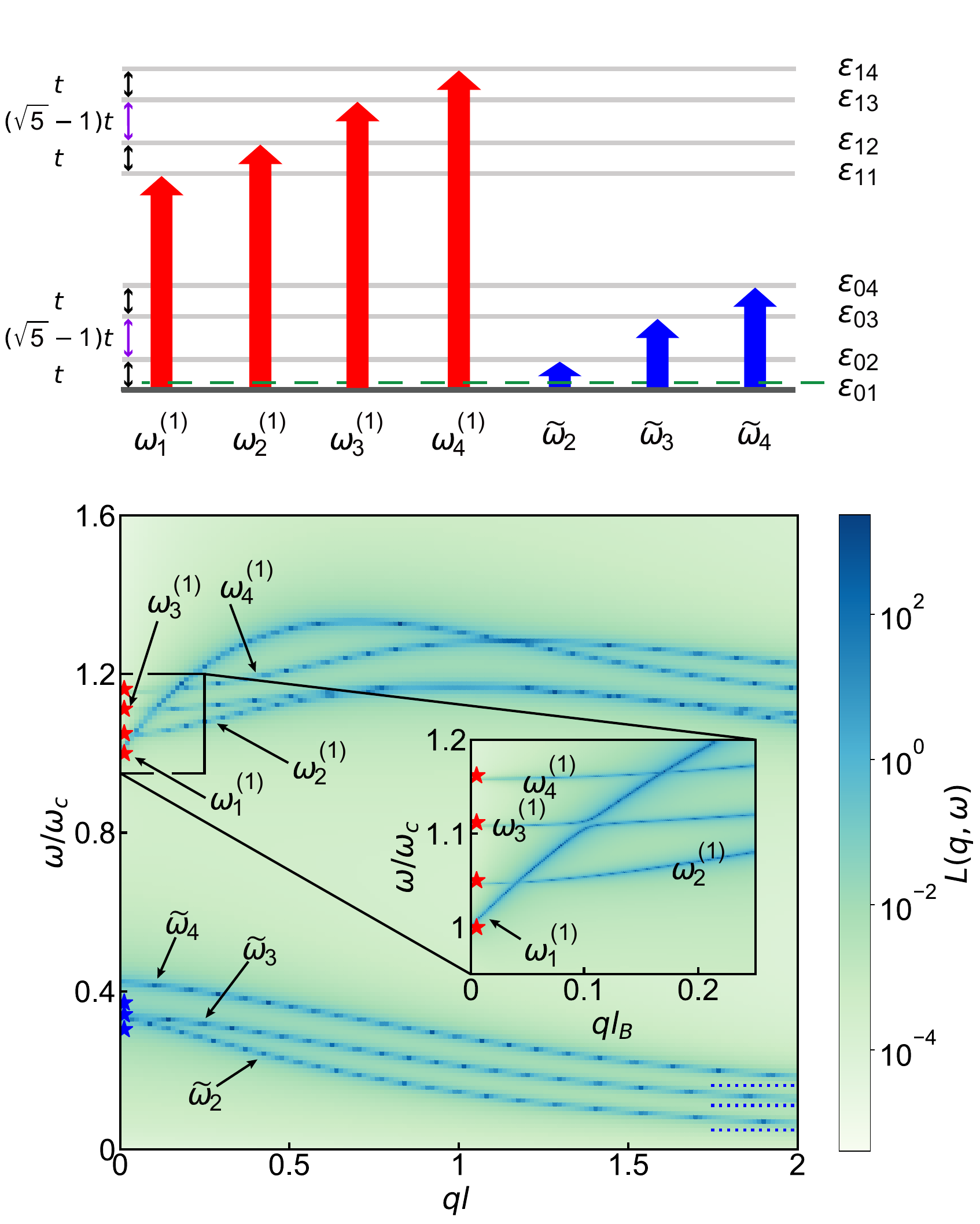} 
    \caption{
Schematic illustration of magnetoplasmons associated with each band transition in the weak Coulomb interaction limit, together with the loss function plot
$L(\mathbf{q},\omega)=-\mathrm{Im}\,\mathrm{Tr}[\epsilon^{-1}(\mathbf{q},\omega)]$,
for a coupled tetralayer system filled up to the state $\varepsilon_{01}$.
The same parameters as in Fig.~\ref{Fig:numerical_results_N3_coupled}(a) are used with $t=0.5\,\rm{meV}$.
Unlike Fig.~\ref{Fig:numerical_results_N3_coupled}(a), the blue stars obtained from Eq.~(\ref{Eq:TMP_mode_gap}) deviate from the actual tunneling magnetoplasmon gaps, reflecting additional contributions from interband transitions other than the dominant one in the weak Coulomb interaction limit.
By contrast, the red stars obtained from Eq.~(\ref{Eq:MP_gap}) correctly reproduce the magnetoplasmon gaps.
}
    \label{Fig:numerical for N=4}
\end{figure}

In a tetralayer system filled up to the lowest band (see Fig.~\ref{Fig:numerical for N=4}),
there are four magnetoplasmon modes $\omega_{\alpha}^{(k)}$ ($\alpha=1,\ldots,4$) emerging from each multiple $k\omega_c$ $(k=1,2,...)$.
The dominant interband transitions underlying these modes occur between $\varepsilon_{01}$ and $\varepsilon_{k\alpha}$, and anticrossing behavior is observed between $\omega_{1}^{(k)}$ and $\omega_{3}^{(k)}$, analogous to the trilayer case.
For tunneling magnetoplasmons, as in the trilayer case, the selection rule ensures that $\tilde{\omega}_3$ originates from a single interband transition showing excellent agreement with Eq.~(\ref{Eq:TMP_mode_gap}), while $\tilde{\omega}_2$ and $\tilde{\omega}_4$ are coupled,  leading to deviations from  Eq.~(\ref{Eq:TMP_mode_gap}).
Nevertheless, the asymptotic behavior described by Eq.~(\ref{Eq:TMP_mode_infty}) remains valid.

When bands other than the lowest one are filled, additional complexity arises from the unequal spacing between split bands in the presence of interlayer tunneling.
As a result, band transitions rarely share the same energy gap unless they are related by subband-index-reversal symmetry.
When degeneracies not protected by the symmetry do occur, hybridization induces a splitting analogous to that observed in trilayer systems.
This behavior is most commonly observed in transitions with gaps of $k\hbar\omega_c$, as the gap of transitions between $\varepsilon_{n\lambda}$ and $\varepsilon_{n'=n\pm k,\lambda'=\lambda}$ is insensitive to the unequal level spacings.
See App.~\ref{App. N4} for tetralayer systems filled up to $\varepsilon_{02}$.

\begin{figure}[htb!] 
    \centering
    \includegraphics[width=1.0\linewidth]{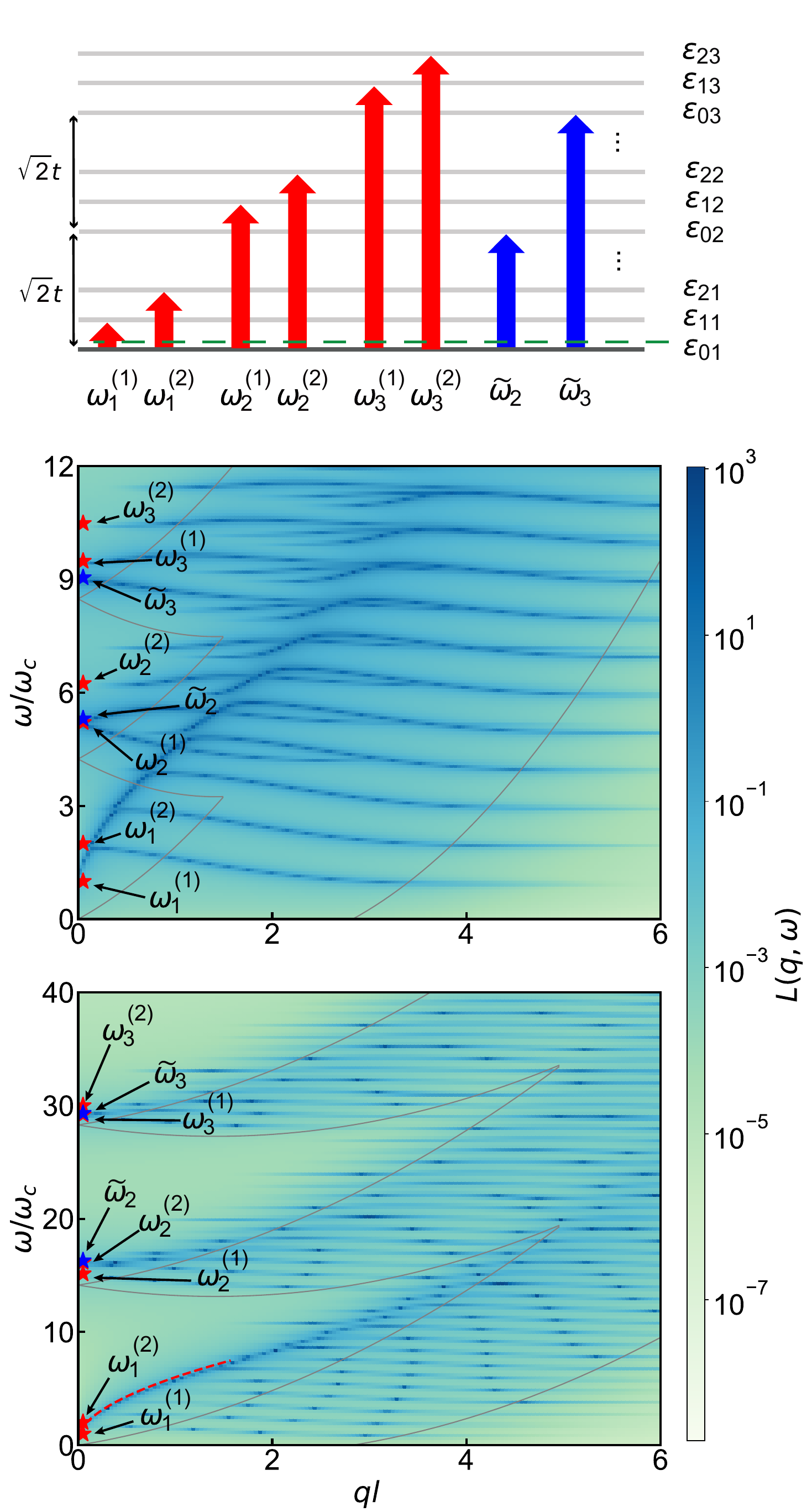} 
    \caption{
The top panel shows a schematic illustration of magnetoplasmons associated with each band transition for a trilayer system filled up to the state $\varepsilon_{01}$ in the limit $t \gg \hbar\omega_c$.
The middle and bottom panels display the loss function plot $L(\mathbf{q},\omega)=-\mathrm{Im}\,\mathrm{Tr}[\epsilon^{-1}(\mathbf{q},\omega)]$ calculated for $t /(\hbar\omega_c)= 3$ and $t /(\hbar\omega_c)= 10$, respectively, with fixed $t = 0.5$ meV.
As in Fig.~\ref{Fig:numerical_results_N3_coupled}, the red and blue stars correspond to results from Eq.~(\ref{Eq:MP_gap}) and Eq.~(\ref{Eq:TMP_mode_gap}), respectively.
The gray lines indicate the boundaries of the electron-hole continuum in a coupled trilayer 2DEG system at zero magnetic field, corresponding to total electron densities of $n_{\mathrm{tot}} = 2.33 \times 10^{9}\,\mathrm{cm}^{-2}$ (middle panel) and $n_{\mathrm{tot}} = 0.70 \times 10^{9}\,\mathrm{cm}^{-2}$ (bottom panel).
}
\label{Fig:B->0 limit}
\end{figure}

When the interlayer tunneling $t$ becomes comparable to $\hbar\omega_c$ (see the middle panel of Fig.~\ref{Fig:B->0 limit}), 
no qualitative differences arise compared to the weak-tunneling regime $t \ll \hbar\omega_c$, except that the ordering of modes in energy can change.
As a consequence, crossing (anticrossing) behavior is observed between magnetoplasmon modes $\omega_{\alpha}^{(k)}$ and tunneling magnetoplasmon modes $\tilde{\omega}_\beta$ with opposite (same) parities of $\alpha$ and $\beta$ when their dispersions meet.
In this regime, Eq.~(\ref{Eq:MP_gap}) and Eq.~(\ref{Eq:MP_gap_modified}) remain valid, and Eq.~(\ref{Eq:TMP_mode_gap}) continues to hold in trilayer systems or in the weak Coulomb interaction limit.
Furthermore, essential features observed when bands other than the lowest one are filled, such as splittings arising from the absence of subband-index-reversal symmetry, persist irrespective of the relative magnitude of $t$ and $\hbar\omega_c$.

In the limit $t \gg \hbar\omega_c$, the dispersion relations in the $B \to 0$ limit are recovered, as the spectral weight distributed among a number of magnetoplasmon modes collectively reconstructs features of non-magnetic systems---specifically, plasmon dispersions and the electron-hole continuum.
Moreover, when the density is confined to the lowest parabolic band of non-magnetic $N$-layer systems, Eq.~(\ref{Eq:TMP}) describes the gap value of out-of-phase plasmon modes~\cite{wvhd-492f} in the $B\to0$ limit, while one in-phase mode is formed by the combination of numerous $\omega_{1}^{(k)}$.
This confirms that tunneling magnetoplasmons are longitudinal modes and originate from the bulk of the $N$-layer system.
See the bottom panel of Fig.~\ref{Fig:B->0 limit} for numerical results in trilayer 2DEG systems with $t/(\hbar\omega_c)=10$ in the regime where the total electron density is confined to the lowest parabolic band.

The theoretical approach developed here is generalizable beyond the trilayer and tetralayer examples discussed. The symmetry-based mode classification offers a robust tool for analyzing complex spectral features in systems with an arbitrary number of layers $N$. Furthermore, we emphasize that the framework can be straightforwardly extended to other multilayer platforms, such as graphene, where unequal Landau level spacing introduces additional complexity. In the case of $N$-layer graphene, the major difference stems from the unequal Landau level spacing, resulting in a complex spectrum of magnetoplasmons compared to that of a 2DEG. However, we observe that our key results can be applied straightforwardly. See App.~\ref{App. graphene} for the numerical results of $N$-layer graphene systems.

In summary, we have established a systematic theoretical framework for analyzing magnetoplasmons in $N$-layer 2DEG systems.
Our analysis clarifies how the magnetoplasmon spectra are governed by the interplay between the Coulomb interaction and interlayer tunneling.
In the decoupled limit, we showed that the single-layer modes split into $N$ branches, whose asymptotic behaviors are obtained in the long-wavelength limit, as well as in the limit of large layer separation and strong magnetic fields.
In the presence of interlayer tunneling, the lifting of Landau level degeneracy allows for multiple interband transitions.
We demonstrated that the spectral structure is determined by the hybridization between magnetoplasmon modes associated with these distinct transitions.
This hybridization is strictly governed by symmetry constraints---specifically parity and subband-index-reversal symmetry---which lead to distinct crossing or anticrossing behaviors.
Furthermore, we identified the origin of tunneling magnetoplasmons arising from intra-Landau-level transitions and provided an analytical description of their gap enhancement in the weak interaction limit. We point out that our approach can be generalized to systems with an arbitrary number of layers $N$ as well as other multilayer platforms including graphene. We hope that our detailed quantitative investigation (comprising both numerical and analytical results) of the magnetoplasmons and loss functions will motivate experimental studies (e.g., involving inelastic light scattering and far-infrared frequency-domain spectroscopy) on collective dynamics in multilayer quantum Hall systems.

\acknowledgments
The work at SNU was supported by the National Research Foundation of Korea (NRF) grants funded by the Korean government (MSIT) (Grant No. RS-2023-NR076715), the Creative-Pioneering Researchers Program through Seoul National University (SNU), and the Center for Theoretical Physics.
EHH acknowledges support from the National Research Foundation of Korea (NRF) (Grant No. RS-2021-NR058646).

\appendix
\section{Eigenvalues and eigenvectors of KMS matrix} \label{App.KMS}
In this section, we review the properties of the eigenvalues and eigenvectors of the KMS matrix defined by 
$A_{ij} = \rho^{|i-j|}$ for $0<\rho<1$ and $1\le i,j \le N$:
\begin{equation}
A =
\begin{pmatrix}
1      & \rho   & \rho^2 & \cdots & \rho^{N-2} & \rho^{N-1} \\
\rho   & 1      & \rho   & \cdots & \rho^{N-3} & \rho^{N-2} \\
\rho^2 & \rho   & 1      & \cdots & \rho^{N-4} & \rho^{N-3} \\
\vdots & \vdots & \vdots & \ddots & \vdots     & \vdots     \\
\rho^{N-2} & \rho^{N-3} & \rho^{N-4} & \cdots & 1 & \rho \\
\rho^{N-1} & \rho^{N-2} & \rho^{N-3} & \cdots & \rho & 1
\end{pmatrix}.
\label{Eq:KMS_matrix}
\end{equation}

The analysis presented below follows Ref.~\cite{wvhd-492f} and its SM.
Here, we omit the explicit derivation and instead focus on extracting the analytical properties relevant to the regime of interest.
The eigenvalues $g_\alpha(\rho)$ and the eigenvectors $\mathbf{u}_\alpha = (u_\alpha^{(1)}, u_\alpha^{(2)},..., u_\alpha^{(N)})^{\mathsf T}$ are given by~\cite{Kac1953, Trench2001, Bogoya2016, Fikioris2018, Narayan2021, wvhd-492f}
\begin{align}
g_{\alpha}(\rho)
&= \frac{1-\rho^{2}}{1-2\rho\cos(\theta_{\alpha})+\rho^{2}},
\label{Eq:g_alpha} \\
u_{\alpha}^{(j)}(\rho)
&= \sin\!\left[
\frac{\alpha}{N+1}j\pi
+ \left(\frac{1}{2}-\frac{j}{N+1}\right)\eta_{\alpha}(\rho)
\right].
\label{Eq:u_alpha_k}
\end{align}
where 
\[
\eta_\alpha(\rho) = 2\arctan\left[\frac{\rho \sin(\theta_\alpha)}{1-\cos(\theta_\alpha)}\right]
\]
and $\theta_\alpha \left(\frac{\alpha-1}{N}\pi <\theta_\alpha<\frac{\alpha}{N}\pi\right)$
is the unique solution to the equation
\begin{equation}
    (N+1)\theta_\alpha+\eta_\alpha(\rho)=\alpha \pi.
\end{equation}

In the limit $\rho\to1$, we find $\eta_\alpha(\rho) \to \pi-\theta_\alpha$ from the relation $\frac{\sin(x)}{1-\cos(x)}=\cot(x/2)$, yielding $\theta_\alpha \to \frac{\alpha-1}{N}\pi$.
The corresponding eigenpairs become
\begin{align}
g_{\alpha}(\rho \to 1)
&=
\begin{cases}
N, & \alpha = 1, \\
\dfrac{1-\rho}{1-\cos\!\left(\dfrac{\alpha-1}{N}\pi\right)}, 
& 2 \le \alpha \le N ,
\end{cases}
\label{AEq:g_alpha_rho1}
\\
u_{\alpha}^{(j)}(\rho \to 1)
&=
\cos\!\left[
\dfrac{(2j-1)(\alpha-1)}{2N}\,\pi
\right].
\label{AEq:u_alpha_k_rho1}
\end{align}

In the limit $\rho\to0$, we have $\eta_\alpha(\rho) \to 0$ and $\theta_\alpha \to \frac{\alpha\pi}{N+1}$, where the eigenpairs take the form
\begin{align}
g_{\alpha}(\rho \to 0)
&= 1+ 2\rho\cos\left(\frac{\alpha \pi}{N+1}\right) + \mathcal{O}(\rho^2),
\label{AEq:g_alpha_rho0}
\\
u_{\alpha}^{(j)}(\rho \to 0)
&=
\sin\!\left(
\dfrac{\alpha j\pi}{N+1}
\right) +\mathcal{O}(\rho).
\label{AEq:u_alpha_k_rho0}
\end{align}
By identifying $\rho=e^{-qd}$, the $\rho\to1$ and $\rho\to0$ limits correspond to the $qd\to0$ and $qd\to\infty$ regimes respectively, and the resulting asymptotic forms for $g_\alpha(e^{-qd})$ and $u_\alpha^{(j)}(e^{-qd})$ are summarized in Table~\ref{Tab:KMS_asymptotics}.


\section{Magnetoplasmon dispersions in single-layer and decoupled $N$-layer 2DEG systems}\label{App.derivation in decoupled limit}
At $T=0$, the noninteracting density-density response function of a single-layer 2DEG at an integer filling factor $\nu$ in a perpendicular magnetic field is given by \cite{Giuliani_Vignale_2005}
\begin{align}
\chi^{(0)}_{\mathrm{2D}}(\mathbf{q},\omega)
= N_0\,\sum_{k=0}^{\infty}
\bar{\chi}^{(0)}_k\left(\bar{q}, \bar{\omega}\right),
\label{AEq:app-chi0}
\end{align}
where $\bar q\equiv \frac{q^2 l^2}{2}$, $\bar\omega\equiv \frac{\omega^+}{\omega_c}$, $\omega^+ \equiv \omega+i0^{+}$ and $ N_0 \equiv \frac{g_s}{2\pi l^2\hbar\omega_c}
= \frac{g_s m}{2\pi\hbar^2}$
is the zero-field density of states at the Fermi energy with $g_s=2$ being the spin degeneracy. The parameters $l=\sqrt{\hbar c/(eB)}$ and $\omega_c=eB/(mc)$ denote the magnetic length and the cyclotron frequency, respectively.
The dimensionless function $\bar{\chi}^{(0)}_{k}(\bar{q},\bar{\omega})$ is defined as
\begin{equation}
\bar{\chi}^{(0)}_k\left(\bar{q}, \bar{\omega}\right) =
\frac{2k}{\bar{\omega}^2 - k^2} 
\sum_{j=j_k}^{\nu-1} \bar{F}_{j+k,k}(\bar{q}),
\label{AEq: 2DEG dim-less χ0}
\end{equation}
where $j_k \equiv \max(0,\nu-k)$ and
\begin{equation}
  \bar{F}_{n,n'}(\bar{q}) = e^{-\bar{q}}\; \bar{q}^{|n-n'|} \frac{n_<!}{n_>!}
\left[L_{n_<}^{\,|n-n'|}\!\left(\bar{q}\right)\right]^2
\label{AEq:dimless_formfactor}
\end{equation}
is the form factor defined in Eq.~(\ref{Eq: 2DEG form factor}), where $n_>=\max(n,n')$ and $n_<=\min(n,n')$.
The associated Laguerre polynomials $L_a^{\,b}(x)$ satisfy the asymptotic forms:
\begin{equation*}
L_j^{k}(x\!\to\!0)=\binom{j+k}{j},\quad
L_j^{k}(x\!\to\!\infty)\sim(-1)^j\frac{x^j}{j!}.
\label{AEq:Laguerre asymptotics}
\end{equation*}
Substituting these forms into Eq.~(\ref{AEq:app-chi0}), we obtain the asymptotic behavior for small and large $\bar{q}$:
\begin{align}
\bar{\chi}^{(0)}_{k}(\bar{q},\bar{\omega}) \to
\begin{cases}
\displaystyle A_k \bar{q}^k \frac{2k}{\bar{\omega}^2 - k^2}, & (\bar{q} \to 0) \\
\displaystyle \frac{e^{-\bar{q}} \bar{q}^{k+2\nu-2}}{(\nu+k-1)!(\nu-1)!} \frac{2k}{\bar{\omega}^2 - k^2}, & (\bar{q} \to \infty)
\end{cases}
\label{AEq: dim-less χ0 q->0}
\end{align}
where $A_k \equiv \sum_{j=j_k}^{\nu-1} \frac{(j+k)!}{j!(k!)^2}$.

The plasmon dispersions are determined within the random phase approximation (RPA) by solving $\epsilon_{\text{RPA}}=0$, where
\begin{align}
\epsilon_{\text{RPA}}(\mathbf{q},\omega)&=1 - v(q)\chi^{(0)}_\text{2D}(\mathbf{q},\omega) \notag \\
&=1 - \frac{g_sme^2\,l}{\kappa \hbar^2}
\frac{1}{\sqrt{2\bar{q}}} \sum_{k=0}^{\infty}\bar{\chi}^{(0)}_k\left(\bar{q}, \bar{\omega}\right)
\label{AEq:single_RPA}
\end{align}
with $v(q)=2\pi e^2/(\kappa q)$ the Coulomb potential and $\kappa$ the dielectric constant.
Near the $k$th cyclotron harmonic ($\omega \simeq k\omega_c$), the response function $\chi^{(0)}_{\text{2D}}$ is dominated by the $k$th term $N_0\bar{\chi}^{(0)}_{k}$ as the denominator $\bar{\omega}^2 - k^2$ approaches zero. 
Accordingly, the dispersion analysis can be simplified by retaining only the $\bar{\chi}^{(0)}_{k}$ term in Eq.~(\ref{AEq:single_RPA}).
This approximation is valid in both the $\bar q\to0$ and $\bar q \to \infty$ limits, and its validity range broadens as the ratio $(me^2\,l)/(\kappa \hbar^2) =(e^2/\kappa l)/(\hbar\omega_c)$ decreases~\cite{Giuliani_Vignale_2005}.
In this regime, the dimensionless asymptotics for \(\bar{\omega}(\bar{q})\) are given by:
\begin{align}
\bar\omega^2 - k^2
&\xrightarrow[\bar q\to 0]{} \sqrt{2}\,k\,A_k\frac{g_sme^2\,l}{\kappa \hbar^2}\,\,\bar q^{\,k-\tfrac12},\\
\bar\omega^2 - k^2 
&\xrightarrow[\bar q\to \infty]{}
\sqrt{2}\,k\,\frac{g_sme^2\,l}{\kappa \hbar^2}
\frac{e^{-\bar q}\,\bar q^{\,k+2\nu-\tfrac52}}{(\nu+k-1)!(\nu-1)!}.
\end{align}
Restoring the units yields the following dispersion relations:
\begin{align}
[\omega^{(k)}(q)]^2
&\xrightarrow[ql\to 0]{} k^2\omega_c^2
+\frac{\sqrt{2}kA_kg_se^2}{\kappa m l^3}\left(\frac{q^2l^2}{2}\right)^{k-1/2},
\label{AEq:GV-dimful-smallq}\\
[\omega^{(k)}(q)]^2
&\xrightarrow[ql\to \infty]{} k^2\omega_c^2 +\frac{\sqrt{2}kg_se^2}{\kappa m l^3} \notag\\
&\quad\times\,
\frac{e^{-q^2l^2/2}}{(\nu+k-1)!(\nu-1)!}
\left(\frac{q^2l^2}{2}\right)^{k+2\nu-5/2},
\label{AEq:GV-dimful-largeq}
\end{align}
where $\omega^{(k)}$ denotes the magnetoplasmon mode emerging from $k\omega_c$.

The extension to decoupled $N$-layer systems is straightforward. In the basis of Coulomb eigenvectors, the RPA dielectric tensor is diagonal, $\epsilon_{\alpha\beta}(\mathbf{q},\omega)=\epsilon_{\alpha}(\mathbf{q},\omega)\,\delta_{\alpha\beta}$, with
\begin{equation} \label{Eq:RPA_decoupled}
    \epsilon_{\alpha} (\mathbf{q},\omega)
    = 1 - g_{\alpha}(e^{-qd})v(q)\chi_{\text{2D}}^{(0)}.
\end{equation}
The single-layer magnetoplasmon mode $\omega^{(k)}$ splits into \(N\) branches $\omega^{(k)}_\alpha$ in the $N$-layer case.
The asymptotics of these modes follow the same functional forms as in Eq.~(\ref{Eq:small_q_chg}) and Eq.~(\ref{Eq:large_q_chg}), obtained by retaining only the contribution from $\bar{\chi}^{(0)}_{k}$ in $\chi^{(0)}_\text{2D}$ when solving the dispersion relation  $\epsilon_\alpha(\mathbf{q},\omega)=0$ and employing the limiting forms of $g_{\alpha}$ (see App.~\ref{App.KMS}).


\section{Tunneling magnetoplasmon dispersions in $N$-layer 2DEG systems filled up to the lowest band}
In this section, we derive Eq.~(\ref{Eq:TMP}) in the regime where the interlayer tunneling $t$ satisfies the weak Coulomb interaction limit, defined by the condition $(e^2/\kappa l)/t\ll1$.
To briefly outline the derivation, we first demonstrate that considering only intra-Landau-level transitions in Eq.~(\ref{Eq: χ0_layerbasis}) is sufficient to determine the gap of the tunneling magnetoplasmon modes.
This step also accounts for the absence of any additional gap enhancement relative to the underlying interband gap in the magnetoplasmon modes $\omega_{\alpha}^{(k)}$, as stated in Eq.~(\ref{Eq:MP_gap}) and Eq.~(\ref{Eq:MP_gap_modified}).
We then derive Eq.~(\ref{Eq:TMP}) by considering only a specific interband transition, which is valid in the weak Coulomb interaction regime owing to the small gap enhancement. 

Analogous to Eq.~(\ref{AEq:app-chi0}), the dimensionless density-density response function for an $N$-layer 2DEG at $T=0$ in the Coulomb eigenbasis, obtained by replacing the indices $ij$ in Eq.~(\ref{Eq: χ0_layerbasis}) with $\alpha\beta$, is given by
\begin{equation}
    \chi^{(0)}_{\alpha\beta}(\mathbf{q}, \omega)=
    N_0 \sum_{\substack{\varepsilon_{n\lambda}>\varepsilon_F \\ \varepsilon_{n'\lambda'}<\varepsilon_F}} \bar{\chi}^{\alpha\beta}_{n\lambda;n'\lambda'} (\bar{q},\bar{\omega}),
    \label{AEq:dimless_chi0_Coulomb_basis}
\end{equation}
where $\varepsilon_{n\lambda}=(n+\frac{1}{2})\hbar\omega_c +\Delta_{\lambda}$ is the energy associated with Landau level index $n$ and subband index $\lambda=1, \dots,N$.
$\Delta_\lambda=-2t\cos\left(\frac{\lambda\pi}{N+1}\right)$ denotes the subband splitting due to interlayer tunneling $t$.
The summation is taken over all $n,\,n',\,\lambda,$ and $\lambda'$ satisfying
$\varepsilon_{n\lambda}>\varepsilon_F$ and $\varepsilon_{n'\lambda'}<\varepsilon_F$.
The dimensionless function $\bar{\chi}^{\alpha\beta}_{n\lambda;n'\lambda'}$ is defined as
\begin{equation}
\bar{\chi}^{\alpha\beta}_{n\lambda;n'\lambda'}(\bar{q},\bar{\omega})\equiv 
\frac{2[(n-n') + \bar{\Delta}_{\lambda\lambda'}]}{\bar{\omega}^2-[(n-n') + \bar{\Delta}_{\lambda\lambda'}]^2} \bar{F}_{n,n'}(\bar{q})P_{\lambda\lambda'}^{\alpha\beta}.     
\end{equation}
where \( \bar{F}_{n,n'}(\bar{q})\) is the form factor defined in Eq.~(\ref{AEq:dimless_formfactor}) and
$P^{\lambda\lambda'}_{\alpha\beta}
= \langle \lambda | U_\alpha | \lambda' \rangle
\langle \lambda' | U_\beta | \lambda \rangle$ is the overlap factor defined in Eq.~(\ref{Eq: overlap_Coulomb}).

The determination of the dispersion relations by solving $\det[\epsilon_{\alpha\beta}(\mathbf{q},\omega)]=0$ is more intricate than the analysis presented in App.~\ref{App.derivation in decoupled limit}.
The dielectric matrix is given by 
\begin{align}
\epsilon_{\alpha\beta}(\mathbf{q},\omega)
&= \delta_{\alpha\beta} - g_\alpha(e^{-qd})v(q)\,\chi^{(0)}_{\alpha\beta}(\mathbf{q},\omega) \notag\\
&=\delta_{\alpha\beta} - \frac{gme^2\,l}{\kappa \hbar^2}
\frac{g_\alpha(e^{-qd})}{\sqrt{2\bar{q}}}
\sum_{\substack{\varepsilon_{n\lambda}>\varepsilon_F\\\varepsilon_{n'\lambda'}<\varepsilon_F}}\bar{\chi}^{\alpha\beta}_{n\lambda;n'\lambda'}(\bar{q},\bar{\omega})
\label{AEq:epsilon_Coulomb_basis}
\end{align}
in agreement with Eq.~(\ref{Eq:epsilon_Coulomb_basis}).
Notably, in the limit $\bar{q}\to0$, the form factor $\bar{F}_{n,n'}(\bar{q})$ vanishes for $n\neq n'$ due to the $\bar{q}^{|n-n'|}$ factor, which facilitates a tractable analysis of the gap values for each mode.
Specifically, in the limit $\bar{q}\to0$, $g_\alpha(e^{-qd})$ scales as $qd\sim\sqrt{\bar{q}}$ for $\alpha\neq1$ and remains constant for $\alpha=1$ (see Table~\ref{Tab:KMS_asymptotics}).
In either case, $\frac{g_\alpha(e^{-qd})}{\sqrt{\bar q}}\bar{\chi}^{\alpha\beta}_{n\lambda;n'\neq n,\lambda'}$ vanishes unless the denominator $\bar{\omega}^2-[(n-n')+\bar{\Delta}_{\lambda\lambda'}]^2$ approaches zero.
In contrast, the terms with \(n'=n\) can yield a finite contribution to the total response function.

To determine the solutions of $\det(\epsilon_{\alpha\beta})=0$ apart from $[\bar{\omega}(\bar{q}\to0)]^2=[(n-n')+\bar{\Delta}_{\lambda\lambda'}]^2$, it is sufficient to consider only the contribution from $\bar{\chi}^{\alpha\beta}_{n\lambda;n'=n,\lambda'}$ in Eq.~(\ref{AEq:epsilon_Coulomb_basis}).
Physically, this corresponds to transitions between bands with the same Landau level index, giving rise to tunneling magnetoplasmon modes.
On the other hand, magnetoplasmons emerge from the solutions $[\bar{\omega}(\bar{q}\to0)]^2=[(n-n')+\bar{\Delta}_{\lambda\lambda'}]^2$, analogous to the analysis in App.~\ref{App.derivation in decoupled limit}. No additional finite Coulomb correction occurs relative to the interband gap, as stated in Eq.~(\ref{Eq:MP_gap}) and Eq.~(\ref{Eq:MP_gap_modified}).

When the system is filled up to the lowest band $\varepsilon_{01}$, restricting the summation to intra-Landau-level transitions allows
Eq.~(\ref{AEq:dimless_chi0_Coulomb_basis}) to be simplified as
\begin{equation}\label{AEq:chi0_Coulomb_basis_TM}
    \chi_{\alpha\beta}^{(0)}(ql\to0,\omega) \approx \sum_{\lambda=2}^{N}
    P_{\alpha\beta}^{\lambda1} \frac{g}{2\pi l^2} \frac{2\Delta_{\lambda1}}{\hbar^2\omega^2-\Delta_{\lambda1}^2},
\end{equation}
which is formally equivalent to the situation discussed in Sec.~\RNum{4} of the SM in Ref.~\cite{wvhd-492f}.
It is worth emphasizing that the key results of Sec.~\RNum{3} of the SM in Ref.~\cite{wvhd-492f} remain valid in our system owing to the identical structure of the overlap factor $P_{\alpha\beta}^{\lambda\lambda'}$, which is denoted as $F_{\alpha\beta}^{\lambda\lambda'}$ in the cited work.

The final step of the derivation consists of retaining only the term proportional to $1/(\hbar^2\omega^2-\Delta_{\alpha1}^2)$, which yields Eq.~(\ref{Eq:TMP_mode_gap}) following Eq.~(S23) in the cited work.
This procedure is justified under the condition $(e^2/\kappa l)/t \ll 1$, since in this regime, the Coulomb correction term in Eq.~(\ref{Eq:TMP_mode_gap})---given explicitly as
\[
(V_\alpha P_{\alpha,\alpha}^{\alpha1} + V_{\alpha+2} P_{\alpha+2,\alpha+2}^{\alpha1}) \frac{g\Delta_{\alpha1}}{\pi l^2} \sim \frac{e^2}{\kappa lt} 
\]
---becomes negligible.
Consequently, the term containing $1/(\hbar^2\omega^2-\Delta_{\alpha1}^2)$ dominates Eq.~(\ref{AEq:chi0_Coulomb_basis_TM}), justifying the single-term consideration.

In the limit $\bar{q}\to \infty$, the form factor $\bar{F}_{n,n'}(\bar{q})$ vanishes irrespective of the values of $n$ and $n'$; thus, Eq.~(\ref{Eq:TMP_mode_infty}) remains valid by arguments analogous to those for Eq.~(\ref{Eq:MP_gap}) and Eq.~(\ref{Eq:MP_gap_modified}). 


\section{Magnetoplasmons in a tetralayer 2DEG system filled up to $\varepsilon_{02}$} \label{App. N4}

\begin{figure}
    \centering
    \includegraphics[width=1.0\linewidth]{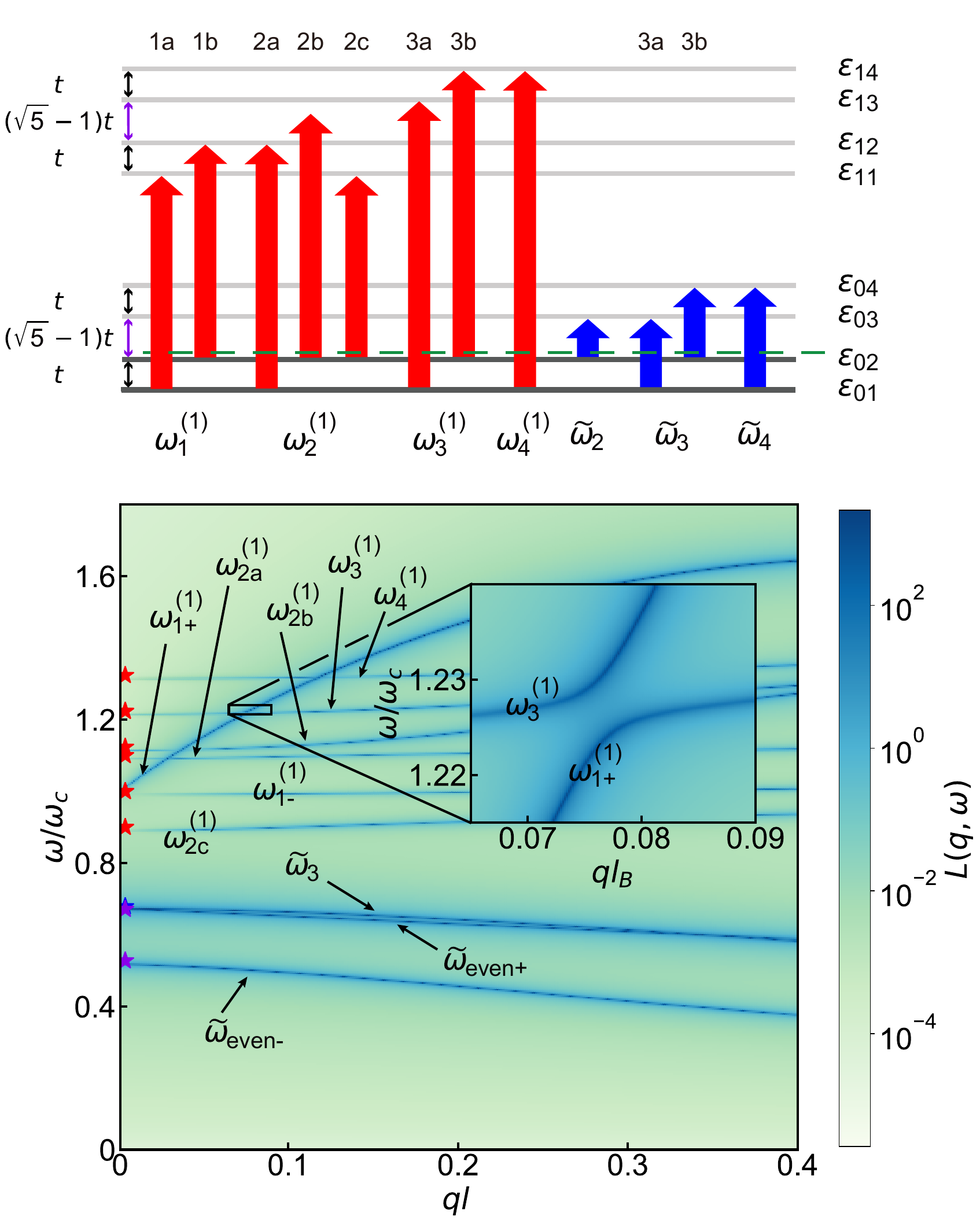} 
    \caption{
Schematic illustration of magnetoplasmons associated with each band transition in the weak Coulomb interaction limit, together with the loss function plot
$L(\mathbf{q},\omega)=-\mathrm{Im}\,\mathrm{Tr}[\epsilon^{-1}(\mathbf{q},\omega)]$,
for a coupled tetralayer system filled up to the state $\varepsilon_{02}$.
The same parameters as in Fig.~\ref{Fig:numerical_results_N3_coupled}(a) are used with $\hbar\omega_c =5 \,\rm{meV}$ and $t=0.5\,\rm{meV}$.
The classification between tunneling magnetoplasmon modes is possible owing to the blue star obtained from Eq.~(\ref{AEq:TMP_mode_gap_tetra_3}) and the purple stars obtained from  and Eq.~(\ref{AEq:TMP_mode_gap_tetra_even}).
As in Fig.~\ref{Fig:numerical for N=4}, the red stars obtained from Eq.~(\ref{Eq:MP_gap_modified}) correctly reproduce the magnetoplasmon gaps.
}
\label{Fig:numeric_N=4_lambda=2}
\end{figure}

In this section, we consider a tetralayer 2DEG system filled up to $\varepsilon_{02}$ (see Fig.~\ref{Fig:numeric_N=4_lambda=2}).
The eight interband transitions between $\varepsilon_{0\lambda}$ and $\varepsilon_{1\lambda'}$ and the corresponding magnetoplasmon modes can be grouped in the same manner as the example illustrated in Fig.~\ref{Fig:numerical_results_N3_coupled}(b).
The hybridization rules, which are governed by parity and subband-index-reversal symmetries, remain equally valid here. 
For example, $\omega^{(1)}_{1a}$ and $\omega^{(1)}_{1b}$ hybridize to form $\omega^{(1)}_{1\pm}$.
The modes $\omega^{(1)}_{3a}$ and $\omega^{(1)}_{3b}$ are related by subband-index-reversal symmetry; therefore, no splitting occurs and a single mode $\omega^{(1)}_{3}$ emerges at $\omega_c+\sqrt{5}t/\hbar$.  
The band transitions associated with $\omega^{(1)}_{2a}$, $\omega^{(1)}_{2b}$, $\omega^{(1)}_{2c}$, and $\omega^{(1)}_4$ all possess distinct underlying band gaps, and thus each forms a separate mode.
As a result, the eight transitions give rise to a total of seven modes, all of which exhibit gaps consistent with Eq.~(\ref{Eq:MP_gap_modified}).

Similarly to the tetralayer case filled up to the lowest band (see Fig.~\ref{Fig:numerical for N=4}), the tunneling magnetoplasmon modes can exhibit deviations from Eq.~(\ref{Eq:TMP}) due to the coupling between plasmon modes.
Thus, we need to consider a matrix form of $\chi_{\alpha\beta}^{(0)}$, which takes the form
\begin{equation} \label{AEq:chi0_Coulomb_basis_TM_generalized}
       \chi_{\alpha\beta}^{(0)}(ql\to0,\omega) \approx \sum_{\lambda=3}^{4}\sum_{\lambda'=1}^{2}
    P_{\alpha\beta}^{\lambda \lambda'} \frac{g}{2\pi l^2} \frac{2\Delta_{\lambda\lambda'}}{\hbar^2\omega^2-\Delta_{\lambda\lambda'}^2},
\end{equation}
where only intra-Landau-level transitions contribute to $\chi_{\alpha\beta}^{(0)}$.
Consequently, the tunneling magnetoplasmon modes are determined by solving the condition $\det(\epsilon_{\alpha\beta})=0$, incorporating this explicit matrix form of $\chi_{\alpha\beta}^{(0)}$.

The selection rule inherent to the Coulomb eigenbasis simplifies the dispersion analysis.
Analogous to the case of $\omega^{(k)}_\alpha$, there are four interband transitions between $\varepsilon_{0\lambda}$ and $\varepsilon_{0\lambda'}$, and the corresponding tunneling magnetoplasmon modes \(\tilde{\omega}_\alpha\) can be labeled in a similar manner.
$\tilde{\omega}_{3a}$ and $\tilde{\omega}_{3b}$ do not couple to either $\tilde{\omega}_{2}$ or $\tilde{\omega}_{4}$ owing to their opposite parity.
Moreover, since $\tilde{\omega}_{3a}$ and $\tilde{\omega}_{3b}$ are related by subband-index-reversal symmetry, they do not split into two modes but instead form a single mode $\tilde{\omega}_3$ determined by 
\begin{equation} \label{AEq:TMP_mode_gap_tetra_3}
    [\hbar\tilde{\omega}_{3}(ql\to 0)]^2= \Delta_{31}^2 + 2V_3 P_{3,3}^{3 1} \frac{g \Delta_{31}}{\pi l^2},
\end{equation}
where the degeneracy of the two modes leads to a factor of two relative to Eq.~(\ref{Eq:MP_gap}).

The remaining two modes $\tilde{\omega}_{2}$ and $\tilde{\omega}_{4}$ hybridize to form $\tilde{\omega}_{\mathrm{even}\pm}$, determined by the condition
\begin{equation} \label{AEq:TMP_mode_gap_tetra_even}
\begin{aligned}
    &\det [ \epsilon_{\alpha\beta} ( q \to 0, \omega ) ]  \\
    &\to\left[ 1 - V_{2} \chi_{2 2}^{(0)} \right] \left[ 1 - V_{4} \chi_{44}^{(0)} \right] 
- V_{2} V_{4} \left[ \chi_{2 4}^{(0)} \right]^2= 0,
\end{aligned}
\end{equation}
where 
\[
\chi_{\alpha \beta}^{(0)} \to \frac{g}{2\pi l^2}\left( 
 P_{\alpha\beta}^{32}\frac{2\Delta_{32}}{\hbar^2\omega^2-\Delta_{32}^2} +P_{\alpha\beta}^{41}\frac{2\Delta_{41}}{\hbar^2\omega^2-\Delta_{41}^2}
\right)
\]
accounts only for the contributions from the band transitions associated with $\tilde{\omega}_{2}$ and $\tilde{\omega}_{4}$ in Eq.~(\ref{AEq:chi0_Coulomb_basis_TM_generalized}). 
This expression yields a quadratic equation in $\omega^2$ that admits analytical solutions $\tilde{\omega}_{\mathrm{even}\pm}$, as illustrated by the purple stars in Fig.~\ref{Fig:numeric_N=4_lambda=2}.


\section{Magnetoplasmons in $N$-layer graphene systems}  \label{App. graphene}

\begin{figure}[b]
    \centering
    \includegraphics[width=1.0\linewidth]{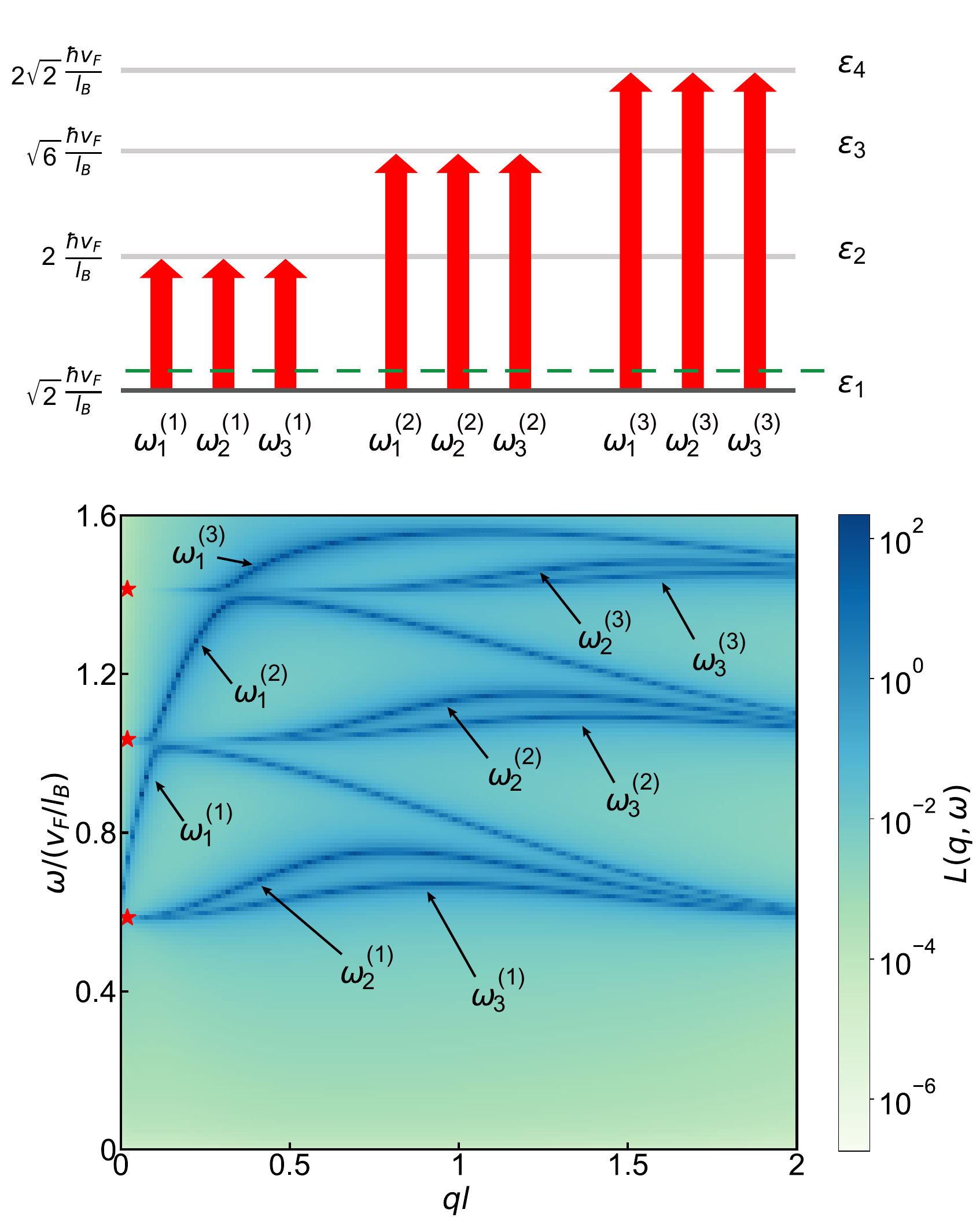} 
    \caption{
    Schematic illustration of magnetoplasmons associated with each band transition, together with the loss function plot
$L(\mathbf{q},\omega)=-\mathrm{Im}\,\mathrm{Tr}[\epsilon^{-1}(\mathbf{q},\omega)]$,
for a decoupled trilayer graphene system filled up to the state $\varepsilon_{1}$.
We use the parameters $\kappa = 3.03$, $d = 20$ \r{A}, $\hbar v_{\rm{F}} = \frac{\sqrt{3}}{2} \gamma_0 a$, $a = 2.46$ \r{A}, $\gamma_0 = 3.1$ eV, $l = 100$ \r{A}, and $\eta = 10^{-3}\,\hbar \omega_{B} = 10^{-3} \sqrt{2} \,\hbar v_{\mathrm{F}} / l$. 
}
\label{Fig:Graphene}
\end{figure}

In this section, we investigate the magnetoplasmon dispersions in $N$-layer graphene systems.
Unlike conventional 2DEGs, graphene exhibits unequally spaced Landau levels with an energy spectrum given by $\varepsilon_{n} = \mathrm{sgn}(n) \frac{\hbar v_F}{l} \sqrt{2|n|}$, where $v_F$ is the Fermi velocity, $l=\sqrt{\hbar c/(eB)}$ is the magnetic length, and the integer $n$ represents an electron-like ($n>0$) or hole-like ($n<0$) Landau level index. 
Moreover, a single Landau level exists for the case $n=0$ with $\varepsilon_0=0$.
The bare susceptibility in single-layer graphene required for Eq.~(\ref{Eq:RPA_decoupled}) was calculated using the graphene form factor in Ref.~\cite{Roldan2009}.
Despite the complexity arising from this unequal Landau level spacing, each dispersion mode $\omega^{(k)}$ in single-layer graphene emerges from the associated inter-Landau-level transitions, analogous to the 2DEG case.
Here we label the magnetoplasmon modes with a positive integer $k$, such that $k=1$ corresponds to the lowest energy mode, $k=2$ to the next lowest, and so forth.
In a decoupled $N$-layer graphene system filled up to the state $\varepsilon_{1}$, the dispersion $\omega^{(k)}$ splits into $\omega^{(k)}_{\alpha}$ $(\alpha=1,\dots,N)$ due to the layer degrees of freedom, similar to the decoupled 2DEG multilayer systems illustrated in Fig.~\ref{Fig: decoupled N=3} (see Fig.~\ref{Fig:Graphene} for trilayer graphene results).
$\omega_{1}^{(k)}$ corresponds to one in-phase mode and $\omega_{\alpha\neq1}^{(k)}$ corresponds to out-of-phase modes.

It is worth noting that, in principle, the hybridization governed by parity and subband-index-reversal symmetry remains valid in the presence of interlayer tunneling $t$.
This is because the subband wave function is separable from the in-plane Landau level wave function [see Eq.~(\ref{Eq:subband wavefunction})], which allows the selection rules inherent in the Coulomb eigenvector basis to apply in the same manner as in the main text.
Similarly, the tunneling magnetoplasmon $\tilde{\omega}_\alpha$ arises from intra-Landau-level transitions, subject to additional Coulomb corrections relative to the underlying interband gap.
However, the application of these rules to clarify the dispersion structures in multilayer graphene is often limited due to the complexity arising from the $t \gg \hbar\omega_c$ condition in general multilayer graphene systems~\cite{Tahir2008, Berman2008, Neto2009} and from the unequally spaced Landau levels.

\vfill

\bibliography{references}

@book{Giuliani_Vignale_2005,
  title = {{Quantum Theory of the Electron Liquid}},
  author = {Giuliani, G. F. and Vignale, G.},
  publisher = {Cambridge University Press},
  year = {2005}
}

@book{mahan1990many,
  title={{Many-Particle Physics}},
  author={Mahan, G.D.},
  isbn={9780306434235},
  lccn={89048850},
  url={https://books.google.co.kr/books?id=v8du6cp0vUAC},
  year={1990},
  publisher={Springer US}
}

@article{Bernstein1958,
  title = {{Waves in a Plasma in a Magnetic Field}},
  author = {Bernstein, Ira B.},
  journal = {Phys. Rev.},
  volume = {109},
  pages = {10},
  year = {1958},
  doi = {10.1103/PhysRev.109.10}
}

@article{Chiu1974,
  title = {{Plasma oscillations of a two-dimensional electron gas in a strong magnetic field}},
  author = {Chiu, K. W. and Quinn, J. J.},
  journal = {Phys. Rev. B},
  volume = {9},
  issue = {11},
  pages = {4724--4732},
  numpages = {0},
  year = {1974},
  month = {Jun},
  publisher = {American Physical Society},
  doi = {10.1103/PhysRevB.9.4724},
  url = {https://link.aps.org/doi/10.1103/PhysRevB.9.4724}
}

@article{Kallin1984,
  title = {{Excitations from a filled {Landau} level in the two-dimensional electron gas}},
  author = {Kallin, C. and Halperin, B. I.},
  journal = {Phys. Rev. B},
  volume = {30},
  pages = {5655},
  year = {1984},
  doi = {10.1103/PhysRevB.30.5655}
}

@article{wvhd-492f,
  title = {{Plasmons in $N$-layer systems}},
  author = {Kim, Taehun and Hwang, E. H. and Min, Hongki},
  journal = {Phys. Rev. B},
  volume = {112},
  issue = {4},
  pages = {L041111},
  numpages = {6},
  year = {2025},
  month = {Jul},
  publisher = {American Physical Society},
  doi = {10.1103/wvhd-492f},
  url = {https://link.aps.org/doi/10.1103/wvhd-492f}
}

@ARTICLE{Kac1953,
    author = "M. Kac, W. Murdock, G. Szego",
     title = {{On the Eigen-Values of Certain Hermitian Forms}},
   journal = "Indiana Univ. Math. J.",
  fjournal = "Indiana University Mathematics Journal",
    volume = 2,
      year = 1953,
     issue = 4,
     pages = "767--800",
      issn = "0022-2518",
     coden = "IUMJAB",
   mrclass = "",
}

@article{Trench2001,
  title = {{Properties of Some Generalizations of {Kac--Murdock--Szeg\"{o}} Matrices}},
  author = {Trench, William F.},
  journal = {Contemp. Math.},
  volume = {281},
  pages = {233},
  year = {2001},
  doi = {10.1090/conm/281/04664}
}

@article{Bogoya2016,
  title = {{Eigenvectors of {Hermitian} Toeplitz matrices with smooth simple-loop symbols}},
  author = {Bogoya, J. M. and B\"{o}ttcher, A. and Grudsky, S. M. and Maximenko, E. A.},
  journal = {Linear Algebra Appl.},
  volume = {493},
  pages = {606},
  year = {2016},
  doi = {10.1016/j.laa.2015.12.017}
}

@article{Fikioris2018,
  title = {{Spectral properties of {Kac--Murdock--Szeg\"{o}} matrices with a complex parameter}},
  author = {Fikioris, George},
  journal = {Linear Algebra Appl.},
  volume = {553},
  pages = {182},
  year = {2018},
  doi = {10.1016/j.laa.2018.05.004}
}

@article{Narayan2021,
  title = {{Generalized {Toeplitz--Hankel} matrices and their application to a layered electron gas}},
  author = {Narayan, Onuttom and Shastry, B. Sriram},
  journal = {J. Phys. A: Math. Theor.},
  volume = {54},
  pages = {175201},
  year = {2021},
  doi = {10.1088/1751-8121/abee67}
}

@article{Roldan2009,
  title = {{Collective modes of doped graphene and a standard two-dimensional electron gas in a strong magnetic field: Linear magnetoplasmons versus magnetoexcitons}},
  author = {Rold\'an, R. and Fuchs, J.-N. and Goerbig, M. O.},
  journal = {Phys. Rev. B},
  volume = {80},
  issue = {8},
  pages = {085408},
  numpages = {6},
  year = {2009},
  month = {Aug},
  publisher = {American Physical Society},
  doi = {10.1103/PhysRevB.80.085408},
  url = {https://link.aps.org/doi/10.1103/PhysRevB.80.085408}
}

@article{DasSarma1982,
  title = {{Collective excitations in semiconductor superlattices}},
  author = {Das Sarma, S. and Quinn, J. J.},
  journal = {Phys. Rev. B},
  volume = {25},
  issue = {12},
  pages = {7603--7618},
  numpages = {0},
  year = {1982},
  month = {Jun},
  publisher = {American Physical Society},
  doi = {10.1103/PhysRevB.25.7603},
  url = {https://link.aps.org/doi/10.1103/PhysRevB.25.7603}
}

@article{Bonanni2011,
  author = {Bonanni, Valentina and Bonetti, Stefano and Pakizeh, Tavakol and Pirzadeh, Zhaleh and Chen, Jianing and Nogués, Josep and Vavassori, Paolo and Hillenbrand, Rainer and Åkerman, Johan and Dmitriev, Alexandre},
  title = {{Designer Magnetoplasmonics with {Nickel} Nanoferromagnets}},
  journal = {Nano Lett.},
  year = {2011},
  volume = {11},
  pages = {5333},
  doi = {10.1021/nl2028443}
}

@article{Melander2012,
  author = {Melander, Emil and Östman, Erik and Keller, Janine and Schmidt, Jan and Papaioannou, Evangelos Th. and Kapaklis, Vassilios and Arnalds, Unnar B. and Sanyal, Biplab and Dmitriev, Alexandre and Hjörvarsson, Björgvin},
  title = {{Influence of the magnetic field on the plasmonic properties of transparent {Ni} anti-dot arrays}},
  journal = {Appl. Phys. Lett.},
  year = {2012},
  volume = {101},
  pages = {063107},
  doi = {10.1063/1.4742931}
}

@article{Pineider2013,
  author = {Pineider, Francesco and Campo, Giulio and Bonanni, Valentina and de Juli{\'a}n Fern{\'a}ndez, C{\'e}sar and Mattei, Giovanni and Caneschi, Andrea and Gatteschi, Dante and Sangregorio, Claudio},
  title = {{Circular Magnetoplasmonic Modes in Gold Nanoparticles}},
  journal = {Nano Lett.},
  year = {2013},
  volume = {13},
  pages = {4785},
  doi = {10.1021/nl402394p}
}

@article{Armelles2013,
  author = {Armelles, Gaspar and Cebollada, Alfonso and Garc{\'i}a-Mart{\'i}n, Antonio and Gonz{\'a}lez, Mar{\'i}a Uju{\'e}},
  title = {{Magnetoplasmonics: Combining Magnetic and Plasmonic Functionalities}},
  journal = {Adv. Optical Mater.},
  year = {2013},
  volume = {1},
  pages = {10},
  doi = {10.1002/adom.201200011}
}

@article{Han2017,
  author = {Han, Bing and Gao, Xiaoqing and Shi, Lin and Zheng, Yonglong and Hou, Ke and Lv, Jiawei and Guo, Jun and Zhang, Wei and Tang, Zhiyong},
  title = {{Geometry-Modulated Magnetoplasmonic Optical Activity of {Au} Nanorod-Based Nanostructures}},
  journal = {Nano Lett.},
  year = {2017},
  volume = {17},
  pages = {6083},
  doi = {10.1021/acs.nanolett.7b02583}
}

@article{Maccaferri2015,
  author = {Maccaferri, Nicol{\`o} and Gregorczyk, Keith E. and de Oliveira, Thales V. A. G. and Kataja, Mikko and van Dijken, Sebastiaan and Pirzadeh, Zhaleh and Dmitriev, Alexandre and {\AA}kerman, Johan and Knez, Mato and Vavassori, Paolo},
  title = {{Ultrasensitive and label-free molecular-level detection enabled by light phase control in magnetoplasmonic nanoantennas}},
  journal = {Nat. Commun.},
  year = {2015},
  volume = {6},
  pages = {6150},
  doi = {10.1038/ncomms7150}
}

@article{Profumo2010,
  title = {{Electron-electron interactions in decoupled graphene layers}},
  author = {Profumo, Rosario E. V. and Polini, Marco and Asgari, Reza and Fazio, Rosario and MacDonald, A. H.},
  journal = {Phys. Rev. B},
  volume = {82},
  pages = {085443},
  year = {2010},
  doi = {10.1103/PhysRevB.82.085443}
}

@article{Jang2015,
  title = {{Stacking dependence of carrier interactions in multilayer graphene systems}},
  author = {Jang, Yunsu and Hwang, E. H. and MacDonald, A. H. and Min, Hongki},
  journal = {Phys. Rev. B},
  volume = {92},
  pages = {041411},
  year = {2015},
  doi = {10.1103/PhysRevB.92.041411}
}

@article{Fei2015,
  title = {{Tunneling Plasmonics in Bilayer Graphene}},
  author = {Fei, Z. and Iwinski, E. G. and Ni, G. X. and Zhang, L. M. and Bao, W. and Rodin, A. S. and Lee, Y. and Wagner, M. and Liu, M. K. and Dai, S. and Goldflam, M. D. and Thiemens, M. and Keilmann, F. and Lau, C. N. and Castro-Neto, A. H. and Fogler, M. M. and Basov, D. N.},
  journal = {Nano Lett.},
  volume = {15},
  pages = {4973},
  year = {2015},
  doi = {10.1021/acs.nanolett.5b00912}
}

@article{Liu2014,
  title = {{Evolution of interlayer coupling in twisted molybdenum disulfide bilayers}},
  author = {Liu, Kaihui and Zhang, Liming and Cao, Ting and Jin, Chenhao and Qiu, Diana and Zhou, Qin and Zettl, Alex and Yang, Peidong and Louie, Steve G. and Wang, Feng},
  journal = {Nat. Commun.},
  volume = {5},
  pages = {4966},
  year = {2014},
  doi = {10.1038/ncomms5966}
}

@article{Novelli2020,
  title = {{Optical and plasmonic properties of twisted bilayer graphene: Impact of interlayer tunneling asymmetry and ground-state charge inhomogeneity}},
  author = {Novelli, Pietro and Torre, Iacopo and Koppens, Frank H. L. and Taddei, Fabio and Polini, Marco},
  journal = {Phys. Rev. B},
  volume = {102},
  pages = {125403},
  year = {2020},
  doi = {10.1103/PhysRevB.102.125403}
}

@article{Shin2023,
  title = {{Electronic structure of biased alternating-twist multilayer graphene}},
  author = {Shin, Kyungjin and Jang, Yunsu and Shin, Jiseon and Jung, Jeil and Min, Hongki},
  journal = {Phys. Rev. B},
  volume = {107},
  pages = {245139},
  year = {2023},
  doi = {10.1103/PhysRevB.107.245139}
}

@article{Burg2017,
  title = {{Coherent Interlayer Tunneling and Negative Differential Resistance with High Current Density in Double Bilayer Graphene--{WSe$_2$} Heterostructures}},
  author = {Burg, G. William and Prasad, Nitin and Fallahazad, Babak and Valsaraj, Amithraj and Kim, Kyounghwan and Taniguchi, Takashi and Watanabe, Kenji and Wang, Qingxiao and Kim, Moon J. and Register, Leonard F. and Tutuc, Emanuel},
  journal = {Nano Lett.},
  volume = {17},
  pages = {3919},
  year = {2017},
  doi = {10.1021/acs.nanolett.7b01505}
}

@article{Nguyen2019,
  title = {{Visualizing electrostatic gating effects in two-dimensional heterostructures}},
  author = {Nguyen, Paul V. and Teutsch, Natalie C. and Wilson, Nathan P. and Kahn, Joshua and Xia, Xue and Graham, Abigail J. and Kandyba, Viktor and Giampietri, Alessio and Barinov, Alexei and Constantinescu, Gabriel C. and Yeung, Nelson and Hine, Nicholas D. M. and Xu, Xiaodong and Cobden, David H. and Wilson, Neil R.},
  journal = {Nature},
  volume = {572},
  pages = {220},
  year = {2019},
  doi = {10.1038/s41586-019-1402-1}
}

@article{Seyoung2011,
  title = {{Coulomb drag of massless fermions in graphene}},
  author = {Kim, Seyoung and Jo, Insun and Nah, Junghyo and Yao, Z. and Banerjee, S. K. and Tutuc, E.},
  journal = {Phys. Rev. B},
  volume = {83},
  issue = {16},
  pages = {161401},
  numpages = {4},
  year = {2011},
  month = {Apr},
  publisher = {American Physical Society},
  doi = {10.1103/PhysRevB.83.161401},
  url = {https://link.aps.org/doi/10.1103/PhysRevB.83.161401}
}

@article{Gorbachev2012,
  author = {Gorbachev, R. V. and Geim, A. K. and Katsnelson, M. I. and Novoselov, K. S. and Tudorovskiy, T. and Grigorieva, I. V. and MacDonald, A. H. and Morozov, S. V. and Watanabe, K. and Taniguchi, T. and Ponomarenko, L. A.},
  title = {{Strong Coulomb drag and broken symmetry in double-layer graphene}},
  journal = {Nat. Phys.},
  year = {2012},
  volume = {8},
  pages = {896},
  doi = {10.1038/nphys2441}
}

@article{Song2012,
  title = {{Energy-Driven Drag at Charge Neutrality in Graphene}},
  author = {Song, Justin C. W. and Levitov, Leonid S.},
  journal = {Phys. Rev. Lett.},
  volume = {109},
  issue = {23},
  pages = {236602},
  numpages = {5},
  year = {2012},
  month = {Dec},
  publisher = {American Physical Society},
  doi = {10.1103/PhysRevLett.109.236602},
  url = {https://link.aps.org/doi/10.1103/PhysRevLett.109.236602}
}

@article{Wang2019,
  author = {Wang, Zefang and Rhodes, Daniel A. and Watanabe, Kenji and Taniguchi, Takashi and Hone, James C. and Shan, Jie and Mak, Kin Fai},
  title = {{Evidence of high-temperature exciton condensation in two-dimensional atomic double layers}},
  journal = {Nature},
  year = {2019},
  volume = {574},
  pages = {76},
  doi = {10.1038/s41586-019-1591-7}
}

@article{Wang2020,
  author = {Wang, Lei and Shih, En-Min and Ghiotto, Augusto and Xian, Lede and Rhodes, Daniel A. and Tan, Cheng and Claassen, Martin and Kennes, Dante M. and Bai, Yusong and Kim, Bumho and Watanabe, Kenji and Taniguchi, Takashi and Zhu, Xiaoyang and Hone, James and Rubio, Angel and Pasupathy, Abhay N. and Dean, Cory R.},
  title ={{Correlated electronic phases in twisted bilayer transition metal dichalcogenides}},
  journal = {Nat. Mater.},
  year = {2020},
  volume = {19},
  pages = {861},
  doi = {10.1038/s41563-020-0708-6}
}

@article{Cao2018,
  author = {Cao, Yuan and Fatemi, Valla and Fang, Shiang and Watanabe, Kenji and Taniguchi, Takashi and Kaxiras, Efthimios and Jarillo-Herrero, Pablo},
  title = {{Unconventional superconductivity in magic-angle graphene superlattices}},
  journal = {Nature},
  year = {2018},
  volume = {556},
  pages = {43},
  doi = {10.1038/nature26160}
}

@article{Liu2017,
  author = {Liu, Xiaomeng and Watanabe, Kenji and Taniguchi, Takashi and Halperin, Bertrand I. and Kim, Philip},
  title = {{Quantum Hall drag of exciton condensate in graphene}},
  journal = {Nat. Phys.},
  year = {2017},
  volume = {13},
  pages = {746},
  doi = {10.1038/nphys4116}
}

@article{Li2017,
  author = {Li, J. I. A. and Taniguchi, T. and Watanabe, K. and Hone, J. and Dean, C. R.},
  title = {{Excitonic superfluid phase in double bilayer graphene}},
  journal = {Nat. Phys.},
  year = {2017},
  volume = {13},
  pages = {751},
  doi = {10.1038/nphys4140}
}

@article{Aizin1995,
  title = {{Tunneling magnetoplasmon excitations in the semiclassical limit and integer quantum Hall regime for double-quantum-well systems}},
  author = {A{\u{\i}}zin, G. R. and Gumbs, Godfrey},
  journal = {Phys. Rev. B},
  volume = {52},
  issue = {3},
  pages = {1890--1904},
  numpages = {0},
  year = {1995},
  month = {Jul},
  publisher = {American Physical Society},
  doi = {10.1103/PhysRevB.52.1890},
  url = {https://link.aps.org/doi/10.1103/PhysRevB.52.1890}
}

@article{Tahir2008,
  title = {{Inter-band magnetoplasmons in mono- and bilayer graphene}},
  author = {Tahir, M. and Sabeeh, K.},
  journal = {J. Phys.: Condens. Matter},
  volume = {20},
  pages = {425202},
  year = {2008},
  doi = {10.1088/0953-8984/20/42/425202}
}

@article{Berman2008,
  title = {{Magnetoplasmons in layered graphene structures}},
  author = {Berman, Oleg L. and Gumbs, Godfrey and Lozovik, Yurii E.},
  journal = {Phys. Rev. B},
  volume = {78},
  pages = {085401},
  year = {2008},
  doi = {10.1103/PhysRevB.78.085401}
}

@article{Trang2021,
  author  = {Chi Xuan Trang and Qile Li and Yuefeng Yin and Jinwoong Hwang and Golrokh Akhgar and Iolanda Di Bernardo and Antonija Grubi\v{s}i\'{c}-Cabo and Anton Tadich and Michael S. Fuhrer and Sung-Kwan Mo and Nikhil V. Medhekar and Mark T. Edmonds},
  title   = {{Crossover from 2D Ferromagnetic Insulator to Wide Band Gap Quantum Anomalous Hall Insulator in Ultrathin MnBi2Te4}},
  journal = {ACS Nano},
  year    = {2021},
  volume  = {15},
  pages   = {13444},
  doi     = {10.1021/acsnano.1c03936}
}

@article{Lin2018a,
  author  = {Yu-Chuan Lin and Bhakti Jariwala and Brian M. Bersch and Ke Xu and Sarah M. Eichfeld and Xiaotian Zhang and Tanushree H. Choudhury and Yi Pan and Rafik Addou and Christopher M. Smyth and Jun Li and Kehao Zhang and M. Aman Haque and Stefan F\"{o}lsch and Randall M. Feenstra and Robert M. Wallace and Kyeongjae Cho and Susan K. Fullerton-Shirey and Joan M. Redwing and Joshua A. Robinson},
  title   = {{Realizing Large-Scale, Electronic-Grade Two-Dimensional Semiconductors}},
  journal = {ACS Nano},
  year    = {2018},
  volume  = {12},
  pages   = {965},
  doi     = {10.1021/acsnano.7b07059}
}

@article{Chhowalla2013,
  author  = {Manish Chhowalla and Hyeon Suk Shin and Goki Eda and Lain-Jong Li and Kian Ping Loh and Hua Zhang},
  title   = {{The chemistry of two-dimensional layered transition metal dichalcogenide nanosheets}},
  journal = {Nat. Chem.},
  year    = {2013},
  volume  = {5},
  pages   = {263},
  doi     = {10.1038/nchem.1589}
}

@article{Novoselov2004,
  author  = {K. S. Novoselov and A. K. Geim and S. V. Morozov and D. Jiang and Y. Zhang and S. V. Dubonos and I. V. Grigorieva and A. A. Firsov},
  title   = {{Electric Field Effect in Atomically Thin Carbon Films}},
  journal = {Science},
  year    = {2004},
  volume  = {306},
  pages   = {666},
  doi     = {10.1126/science.1102896}
}

@article{Coleman2011,
  author  = {Jonathan N. Coleman and Mustafa Lotya and Arlene O'Neill and Shane D. Bergin and Paul J. King and Umar Khan and Karen Young and Alexandre Gaucher and Sukanta De and Ronan J. Smith and Igor V. Shvets and Sunil K. Arora and George Stanton and Hye-Young Kim and Kangho Lee and Gyu Tae Kim and Georg S. Duesberg and Toby Hallam and John J. Boland and Jing Jing Wang and John F. Donegan and Jaime C. Grunlan and Gregory Moriarty and Aleksey Shmeliov and Rebecca J. Nicholls and James M. Perkins and Eleanor M. Grieveson and Koenraad Theuwissen and David W. McComb and Peter D. Nellist and Valeria Nicolosi},
  title   = {{Two-Dimensional Nanosheets Produced by Liquid Exfoliation of Layered Materials}},
  journal = {Science},
  year    = {2011},
  volume  = {331},
  pages   = {568},
  doi     = {10.1126/science.1194975}
}

@article{Cho1975,
  author = {Cho, A. Y. and Arthur, J. R.},
  title = {{Molecular beam epitaxy}},
  journal = {Prog. Solid State Chem.},
  year = {1975},
  volume = {10},
  pages = {157},
  doi = {10.1016/0079-6786(75)90008-4}
}

@article{Lei2022,
  author  = {Yu Lei and Tianyi Zhang and Yu-Chuan Lin and Tomotaroh Granzier-Nakajima and George Bepete and Dorota A. Kowalczyk and Zhong Lin and Da Zhou and Thomas F. Schranghamer and Akhil Dodda and Amritanand Sebastian and Yifeng Chen and Yuanyue Liu and Geoffrey Pourtois and Thomas J. Kempa and Bruno Schuler and Mark T. Edmonds and Su Ying Quek and Ursula Wurstbauer and Stephen M. Wu and Nicholas R. Glavin and Saptarshi Das and Saroj Prasad Dash and Joan M. Redwing and Joshua A. Robinson and Mauricio Terrones},
  title   = {{Graphene and Beyond: Recent Advances in Two-Dimensional Materials Synthesis, Properties, and Devices}},
  journal = {ACS Nanosci. Au},
  year    = {2022},
  volume  = {2},
  pages   = {450},
  doi     = {10.1021/acsnanoscienceau.2c00017}
}

@article{Dingle1974,
  title = {{Quantum States of Confined Carriers in Very Thin {${\text{Al}}_{x}{\text{Ga}}_{1\ensuremath{-}x}\text{As}$-GaAs-${\text{Al}}_{x}{\text{Ga}}_{1\ensuremath{-}x}\text{As}$} Heterostructures}},
  author = {Dingle, R. and Wiegmann, W. and Henry, C. H.},
  journal = {Phys. Rev. Lett.},
  volume = {33},
  issue = {14},
  pages = {827--830},
  numpages = {0},
  year = {1974},
  month = {Sep},
  publisher = {American Physical Society},
  doi = {10.1103/PhysRevLett.33.827},
  url = {https://link.aps.org/doi/10.1103/PhysRevLett.33.827}
}

@article{Radisavljevic2011,
  title = {{Single-layer MoS2 transistors}},
  author = {Radisavljevic, B. and Radenovic, A. and Brivio, J. and Giacometti, V. and Kis, A.},
  journal = {Nat. Nanotechnol.},
  volume = {6},
  pages = {147},
  year = {2011},
  doi = {10.1038/nnano.2010.279}
}

@article{Stern1967,
  title = {{Polarizability of a Two-Dimensional Electron Gas}},
  author = {Stern, Frank},
  journal = {Phys. Rev. Lett.},
  volume = {18},
  issue = {14},
  pages = {546--548},
  numpages = {0},
  year = {1967},
  month = {Apr},
  publisher = {American Physical Society},
  doi = {10.1103/PhysRevLett.18.546},
  url = {https://link.aps.org/doi/10.1103/PhysRevLett.18.546}
}

@article{Allen1977,
  title = {{Observation of the Two-Dimensional Plasmon in Silicon Inversion Layers}},
  author = {Allen, S. J. and Tsui, D. C. and Logan, R. A.},
  journal = {Phys. Rev. Lett.},
  volume = {38},
  issue = {17},
  pages = {980--983},
  numpages = {0},
  year = {1977},
  month = {Apr},
  publisher = {American Physical Society},
  doi = {10.1103/PhysRevLett.38.980},
  url = {https://link.aps.org/doi/10.1103/PhysRevLett.38.980}
}

@article{Hwang2007,
  title = {{Dielectric function, screening, and plasmons in two-dimensional graphene}},
  author = {Hwang, E. H. and Das Sarma, S.},
  journal = {Phys. Rev. B},
  volume = {75},
  issue = {20},
  pages = {205418},
  numpages = {6},
  year = {2007},
  month = {May},
  publisher = {American Physical Society},
  doi = {10.1103/PhysRevB.75.205418},
  url = {https://link.aps.org/doi/10.1103/PhysRevB.75.205418}
}

@article{Fei2012,
  author = {Fei, Z. and Rodin, A. S. and Andreev, G. O. and Bao, W. and McLeod, A. S. and Wagner, M. and Zhang, L. M. and Zhao, Z. and Thiemens, M. and Dominguez, G. and Fogler, M. M. and Castro Neto, A. H. and Lau, C. N. and Keilmann, F. and Basov, D. N.},
  title = {{Gate-tuning of graphene plasmons revealed by infrared nano-imaging}},
  journal = {Nature},
  year = {2012},
  volume = {487},
  pages = {82},
  doi = {10.1038/nature11253}
}

@article{Chen2012,
  author = {Chen, J. and Badioli, M. and Alonso-Gonz\'alez, P. and Thongrattanasiri, S. and Huth, F. and Osmond, J. and Spasenovi\'c, M. and Centeno, A. and Pesquera, A. and Godignon, P. and Zurutuza Elorza, A. and Camara, N. and de Abajo, F. J. G. and Hillenbrand, R. and Koppens, F. H. L.},
  title = {{Optical nano-imaging of gate-tunable graphene plasmons}},
  journal = {Nature},
  year = {2012},
  volume = {487},
  pages = {77},
  doi = {10.1038/nature11254}
}

@article{Grigorenko2012,
  author = {Grigorenko, A. N. and Polini, M. and Novoselov, K. S.},
  title ={ {Graphene plasmonics}},
  journal = {Nat. Photonics},
  year = {2012},
  volume = {6},
  pages = {749},
  doi = {10.1038/nphoton.2012.262}
}

@article{Zhang2016,
  author = {Zhang, Qi and Lou, Min and Li, Xinwei and Reno, John L. and Pan, Wei and Watson, John D. and Manfra, Michael J. and Kono, Junichiro},
  title = {{Collective non-perturbative coupling of 2D electrons with high-quality-factor terahertz cavity photons}},
  journal = {Nat. Phys.},
  year = {2016},
  volume = {12},
  pages = {1005},
  doi = {10.1038/nphys3850}
}

@article{Keller2017,
  title={{Few-electron ultrastrong light-matter coupling at 300 GHz with nanogap hybrid LC microcavities}},
  author={Keller, Janine and Scalari, Giacomo and Cibella, Sara and Maissen, Curdin and Appugliese, Felice and Giovine, Ennio and Leoni, Roberto and Beck, Mattias and Faist, J{\'e}r{\^o}me},
  journal={Nano Letters},
  volume={17},
  number={12},
  pages={7410--7415},
  year={2017},
  publisher={ACS}
}

@article{Grigelionis2015,
  title = {{Magnetoplasmons in high electron mobility CdTe/CdMgTe quantum wells}},
  author = {Grigelionis, I. and Nogajewski, K. and Karczewski, G. and Wojtowicz, T. and Czapkiewicz, M. and Wr\'obel, J. and Boukari, H. and Mariette, H. and \L{}usakowski, J.},
  journal = {Phys. Rev. B},
  volume = {91},
  issue = {7},
  pages = {075424},
  numpages = {6},
  year = {2015},
  month = {Feb},
  publisher = {American Physical Society},
  doi = {10.1103/PhysRevB.91.075424},
  url = {https://link.aps.org/doi/10.1103/PhysRevB.91.075424}
}

@article{Lundeberg2017,
  author  = {Lundeberg, Mark B. and Gao, Yuanda and Asgari, Reza and Tan, Cheng and Van Duppen, Ben and Autore, Marta and Alonso-González, Pablo and Woessner, Achim and Watanabe, Kenji and Taniguchi, Takashi and Hillenbrand, Rainer and Hone, James and Polini, Marco and Koppens, Frank H. L.},
  title   = {{Tuning quantum nonlocal effects in graphene plasmonics}},
  journal = {Science},
  volume  = {357},
  number  = {6347},
  pages   = {187--191},
  year    = {2017},
  doi     = {10.1126/science.aan2735}
}

@article{Cao2017,
  author  = {Cao, Lei and Fu, Qiang and Wu, Bang and Xiong, Yongqian},
  title   = {{Terahertz magnetoplasmon-polaritons with nonlocal corrections for lossy two dimensional electron gas in GaN-based heterostructures}},
  journal = {Journal of Physics: Condensed Matter},
  volume  = {29},
  number  = {39},
  pages   = {395302},
  year    = {2017},
  doi     = {10.1088/1361-648X/aa80c6}
}

@article{Neto2009,
  title = {{The electronic properties of graphene}},
  author = {Castro Neto, A. H. and Guinea, F. and Peres, N. M. R. and Novoselov, K. S. and Geim, A. K.},
  journal = {Rev. Mod. Phys.},
  volume = {81},
  issue = {1},
  pages = {109--162},
  numpages = {0},
  year = {2009},
  month = {Jan},
  publisher = {American Physical Society},
  doi = {10.1103/RevModPhys.81.109},
  url = {https://link.aps.org/doi/10.1103/RevModPhys.81.109}
}

\end{document}